\documentclass[aps,prb,reprint,superscriptaddress]{revtex4-2}
\usepackage{graphicx}
\usepackage{amsfonts} 
\usepackage{dcolumn}
\usepackage{bm}
\usepackage{braket}
\usepackage{amsmath}
\usepackage{xurl}
\usepackage[colorlinks,linkcolor=blue,citecolor=blue,urlcolor=blue]{hyperref}
\usepackage{orcidlink}
\begin{document}
	\title{Emergence of Cascading Flat Bands in Breathing Superlattices}
	\affiliation{
		 State Key Laboratory of Semiconductor Physics and Chip Technologies, Institute of Semiconductors, Chinese Academy of Sciences, PO Box 912, Beijing 100083, China
	}
	\affiliation{
		State Key Lab of Fabrication Technologies for Integrated Circuits, Institute of Microelectronics, Chinese Academy of Sciences, Beijing 100029, China
	}

	\affiliation{
		Center for Quantum Matter, Zhejiang University, Hangzhou 310027, China
	}
	\affiliation{
		College of Materials Science and Opto-electronic Technology, University of Chinese Academy of Sciences, Beijing 100049, China
	}

	\author{Moru Song\orcidlink{0009-0003-4842-6959}}
	\affiliation{
		State Key Laboratory of Semiconductor Physics and Chip Technologies, Institute of Semiconductors, Chinese Academy of Sciences, PO Box 912, Beijing 100083, China
	}
	\affiliation{
		College of Materials Science and Opto-electronic Technology, University of Chinese Academy of Sciences, Beijing 100049, China
	}
	\author{Jinyu Hu}
	\affiliation{
		State Key Lab of Fabrication Technologies for Integrated Circuits, Institute of Microelectronics, Chinese Academy of Sciences, Beijing 100029, China
	}
	\author{Lina Shi}
	\email{shilina@ime.ac.cn}
	\affiliation{
		State Key Lab of Fabrication Technologies for Integrated Circuits, Institute of Microelectronics, Chinese Academy of Sciences, Beijing 100029, China
	}

	\author{Yongliang Zhang\orcidlink{0000-0001-9419-6881}}
	\email{ylzhanglight@semi.ac.cn}
	\affiliation{
		State Key Laboratory of Semiconductor Physics and Chip Technologies, Institute of Semiconductors, Chinese Academy of Sciences, PO Box 912, Beijing 100083, China
	}
	\author{Kai Chang\orcidlink{0000-0002-4609-8061}}
	\email{kchang@zju.edu.cn}
	\affiliation{
		Center for Quantum Matter, Zhejiang University, Hangzhou 310027, China
	}
	
	\date{\today}
	\begin{abstract}
	Flat bands have become a pillar of modern condensed matter physics and photonics owing to the vanishing group velocity and diverging density of states.  Here, we present a paradigmatic scheme to construct arbitrary flat bands on demand by introducing a new type breathing superlattice, where both the number and spectral positions of isolated flat bands can be continuously tailored by simply controlling the breathing strength. Microscopically, the momentum-independent interband scatterings near the band edge protect them robust against weak intra-cell disorder. \textcolor{black}{By dimensional reduction, we establish a duality between the one-dimensional (1D) breathing superlattice and the 2D Harper-Hofstadter model, where cascade flat bands naturally emerge as the different orders of Landau levels in the weak magnetic flux limit.} As a proof of concept, photonic flat bands at optical frequencies are experimentally demonstrated with all-dielectric photonic crystal slabs. \textcolor{black}{Finally, we generalize our scheme to 2D systems to realize partial and omnidirectional flat bands, and discuss the achievement of high-quality factors.} Our findings shed new light on the manipulation of flat bands with high band flatness and large usable bandwidth, paving the way for the development of advanced optical devices.
	\end{abstract}
	\maketitle
	\textit{Introduction.—}Band dispersion, essential for understanding the wave dynamics in periodic systems, can be tailored with artificial lattices containing carefully designed building blocks \cite{IBM5391729,PhysRevLett.58.2059,PhysRevLett.58.2486,PhysRevLett.71.2022,science.289.5485.1734}.
	In optics and photonics, engineering the band dispersion is central to mold light propagation \cite{RN590} and tailor light-matter interactions \cite{PhysRevLett.58.2486,RevModPhys.91.025005,RN629}, leading to a variety of novel devices such as microstructured fibers \cite{Markos_2017,10.1126/sciadv.add3522}, metalenses \cite{RN588} and analog optical computating \cite{science.1242818,Guo:18}. Recently, flat bands have attracted significant attention in photonic systems owing to their unique ability to trap light \cite{nanoph-2020-0043,RN560,PhysRevLett.116.066402,PhysRevLett.129.253001,PhysRevLett.132.043803}. The narrow bandwidth of flat bands leads to localization of light with vanishing group velocity and diverging photonic density of states (DOS) \cite{Li:08}, thereby enhancing light-matter interactions for practical applications including lasers \cite{PhysRevLett.132.186902}, sensors \cite{RN577}, \textcolor{black}{filters \cite{RN737},} and  compact free-electron light sources \cite{RN575}. 
	
	In analogy to electronic systems, several systematic methods have been developed to generate flat bands in photonic systems. The standard construction of flat bands requires line-graphs and bipartite lattices with geometric frustrations, such as Kagome \cite{PhysRevB.81.113104,RN502,PhysRevLett.124.183901,Scheer_2023,RN614,danieli2024flat}, Lieb \cite{PhysRevLett.114.245503,RN488,PhysRevB.93.075126,PhysRevB.101.045131,RN614,danieli2024flat} and dice lattices \cite{PhysRevB.84.241103,PhysRevB.103.155155,PhysRevB.103.195442}, where they arise from the unequal number of orbitals in two sublattices \cite{RN537}. Flat bands can also be achieved in non-crystalline systems such as quasi-periodic \cite{RN528} or fractal lattices \cite{RN524}. Very recently, there has been emerging interest in exploiting flat bands in moiré \cite{PhysRevB.82.121407,RN312,RN331,RN441,RN618,RN772} and strained engineering superlattices \cite{PhysRevLett.112.096805, RN522,PhysRevLett.130.216401,RN497,PhysRevB.107.104426,PhysRevResearch.6.013210},  along with their photonic analogs \cite{RN658,RN661,RN311,RN322,PhysRevB.103.214311,RN344,RN660,10.1126/sciadv.adh8498}, by folding and flattening bands through pseudo-magnetic fields. Despite these intensive parallel efforts, however, how to construct arbitrary flat bands on demand remains elusive. For instance, flat bands in bipartite lattices necessitate specific symmetry with nearest hoppings, which are too restrictive for optical systems \cite{RN537}. These flat bands are difficult to isolate from dispersive bands because of the existence of symmetry-protected band touching points \cite{PhysRevB.99.045107,PhysRevB.104.195128}.
	\begin{figure}[b]
		\centering
		\includegraphics[width=0.85\linewidth]{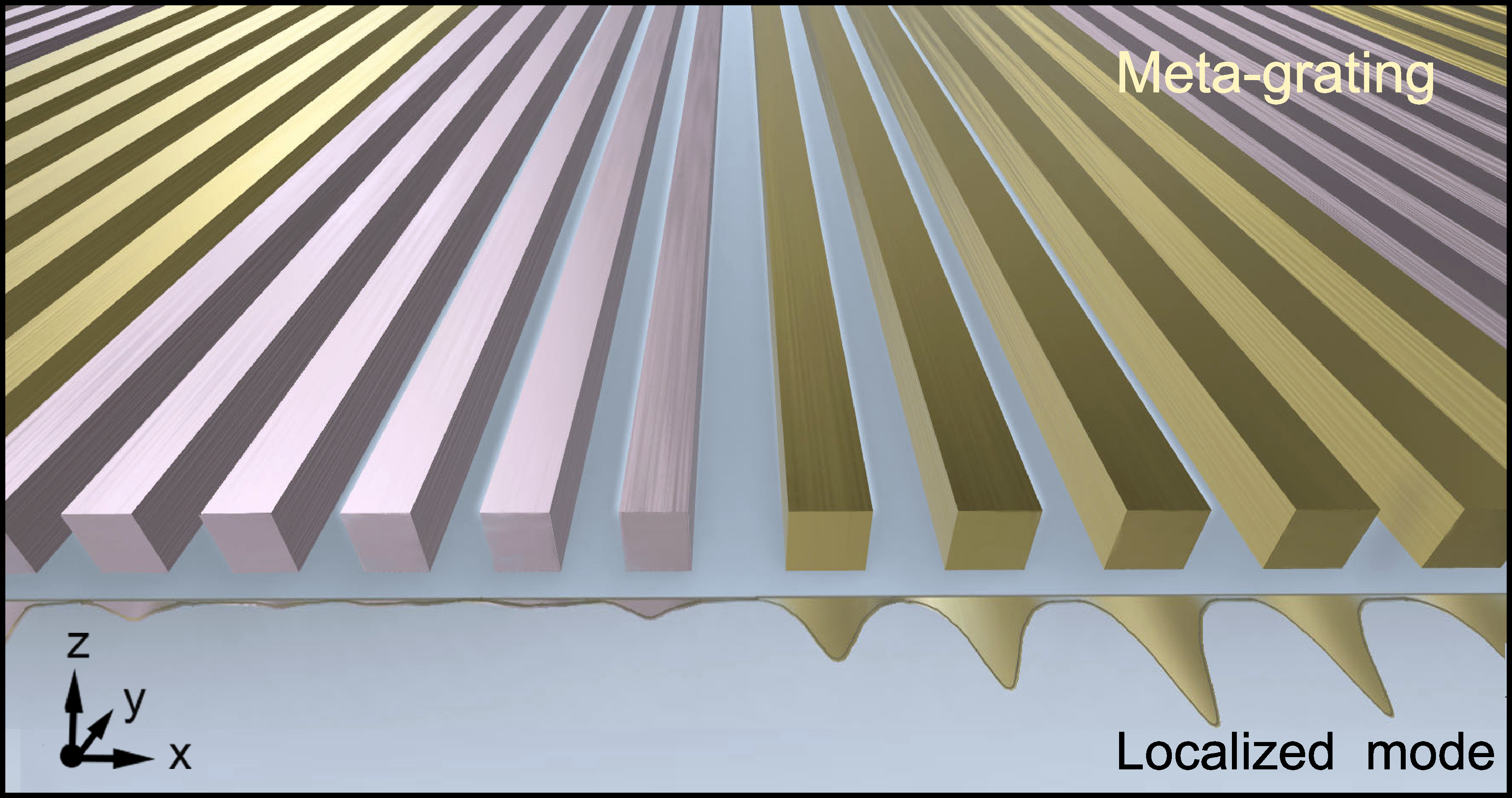}
		\caption{Schematic of the breathing superlattice where the flat band eigenfield is localized in the shrunken sublattice.}
		\label{fig:structurediagram}
	\end{figure}
	\begin{figure*}
		\centering
		\includegraphics[width=0.7\linewidth]{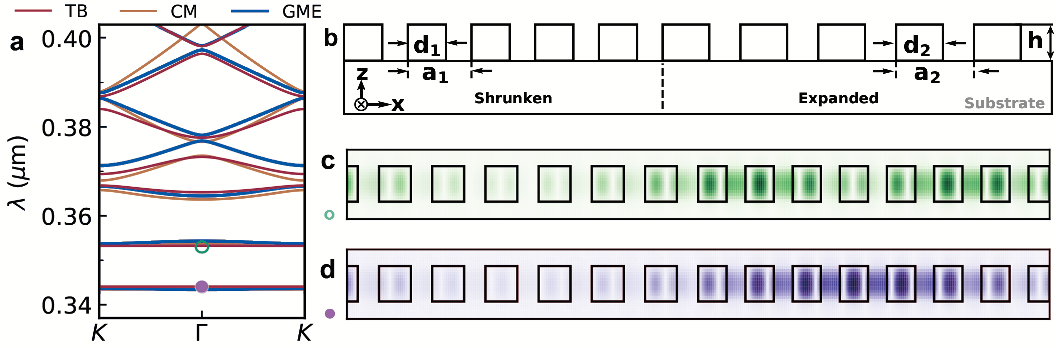}
		\caption { (a) Band structures of a 1D breathing superlattice with $N=14$ for TM polarization, where $K=\pm q_0/2$ and $\Gamma=0$.  The fitting TB parameters are $\mu_i = 730$, $t_0 = 60$ THz, $\delta_T = 0.25$. The fitting CM  parameters are $\mu=850$ , $m^*=1.7$, and $(w_1,w_2,w_3,w_4)=(18,2.88,-4.32, 0.72)$ THz.  (b) Cross-sectional view of the proposed PhCS. (c, d) Electric field distributions at the $\Gamma$ point for higher and lower flat bands.}
		\label{fig:bandstructure}
	\end{figure*}
	To establish destructive interference, non-crystalline systems need specific forms of quasi-perodicity, which hinder the implementation and engineering of flat bands. Moreover, moiré superlattices require precise alignment at the magic angles or distances to create macroscopic moiré patterns \cite{RN312,RN311,RN344}. Most importantly, all known systems possess a fixed number of flat bands, and there has not been a systematic strategy to engineer isolated flat bands with tunable numbers and extreme small bandwidth in a simple platform.
	
	In this Letter, we propose a general approach to construct isolated flat bands on demand by introducing breathing superlattices. Compared with previous studies, both the number and spectral positions of flat bands can be continuously engineered by simply controlling the breathing strength of sublattices. In particular, pairs of isolated flat bands can be cascadingly peeled out from the folded dispersive bands. Analysis based on the continuum model (CM) and the rigorous Maxwell eigenvalue problem demonstrate that these flat bands arise from the interband scatterings near the band edge, which protect them robust against weak intracell disorder. \textcolor{black}{Furthermore, we show that the CM Hamiltonian dual to the Harper-Hofsdter (HH) model \cite{Harper55,SI}, where the cascade flat bands can be interpreted as the Landau levels (LLs) via dimensional reduction \cite{PhysRevB.103.195442}, indicating the topological origin of these flat bands.}  As a concrete example, the formation and evolution of flat bands are experimentally demonstrated with all dielectric photonic crystal slabs. \textcolor{black}{Finally, we extend our approach to two-dimensional systems to enable the realization of both partial and omnidirectional flat bands, and examine the conditions for achieving high quality factors.}
	
	\begin{figure*}[ht]
		\centering
		\includegraphics[width=0.8\linewidth]{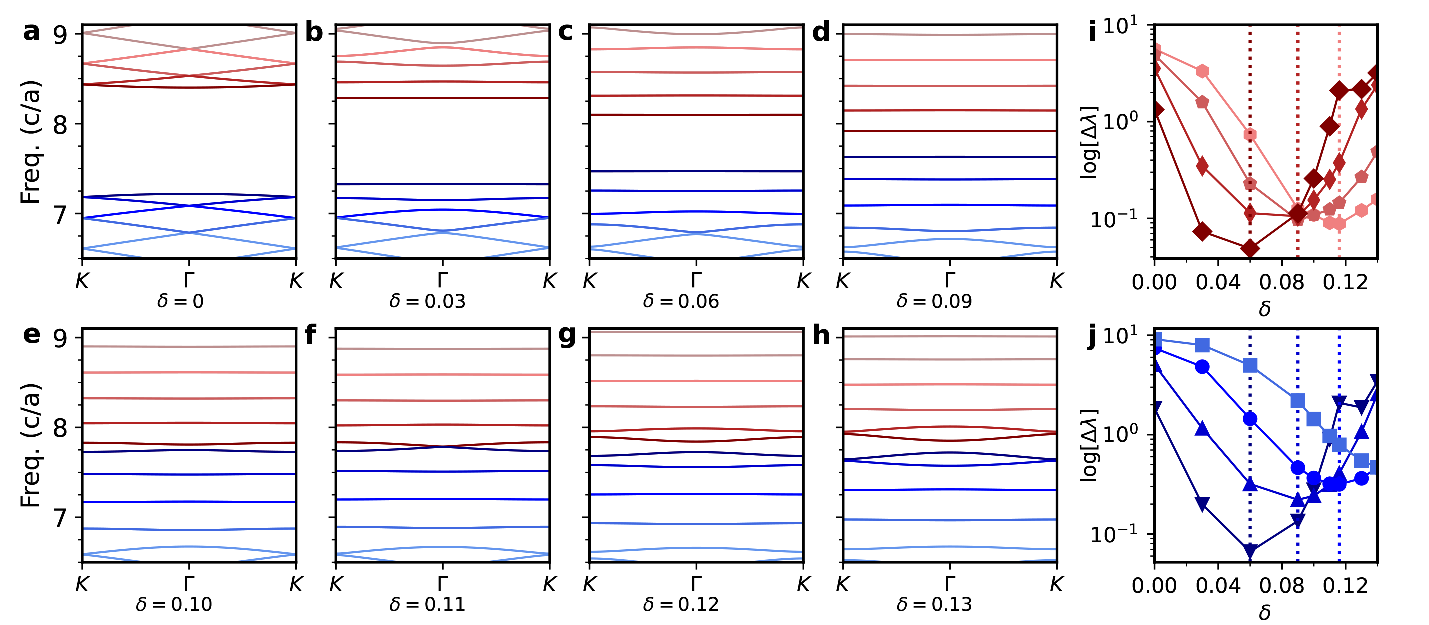}
		\caption{(a-h) Evolution of flat bands for the of PhCSs with $N=14$ and $\delta$ varying from $0$ to $0.14$. Here, the lattice constant is $a=2.66 ~\mu m$. (i,j) Dependence of the bandwidth  ($\Delta\lambda$) of conduction/valence bands on $\delta$. }
		\label{fig:transision}
	\end{figure*}
	
	\textit{Analytical models.—}For simplicity, let us illustrate the scheme by introducing the breathing superlattice in 1D. We start by considering an enlarged unit cell containing $N$ sites in a simple 1D periodic lattice with lattice constant $a_0$. To construct the breathing superlattice, we divide the unit cell into two equal sublattices, and shift them either positively or negatively without altering the size of the large unit cell. Consequently, the unit cell  contains two sublattices, each comprising a simple chain of $n$ sites with different spacings $a_{1,2}=a_0(1\pm \delta)$, where the shifting parameter $\delta$ characterizes the breathing strength. The real-space tight-binding (TB) Hamiltonian takes the form  
	\begin{gather}
		 \hat H = \mu\sum_{Ii}  \hat c_{Ii}^\dagger  \hat c_{Ii} + t\sum_{{Ii}\neq {Jj}} e^{-\xi{|R_{Ii}-R_{Jj}|}} \hat c_{Ii}^\dagger  \hat c_{Jj}, 
		\label{TB_hamiltonian}
	\end{gather}
	where $ \hat c_{Ii}^\dagger$ and $ \hat c_{Ii}$ are creation and annihilation operators at site $i$ of the $I$th unit cell, $\mu$ is the on-site energy, and $te^{-\xi{|R_{Ii}-R_{Jj}|}}$ denotes the hopping between sites $R_{Ii}$ and $R_{Jj}$ with $\xi$ controlling the coupling strength. Note that the structure reduces to the Su–Schrieffer–Heeger model \cite{PhysRevLett.42.1698} when $N=2$ and nearest coupling. For weakly shifted sublattices, the nearest-neighbor hoppings are approximate to $te^{-\xi a_0 (1\pm\delta)}\approx t_0 (1\mp\delta_T)$, where $t_0=te^{-\xi a_0}$ and $\delta_T = \xi a_0 \delta$. By projecting the TB Hamiltonian onto the truncated plane wave basis \cite{RN596}, we obtain the CM Hamiltonian (See Supplementary \cite{SI} S1 for derivations):
 	\begin{gather}
 		 \hat H (q) = \sum_n \left[\frac{(q+Q_n)^2}{2m^*}+ \mu\right] \hat b_n^\dagger  \hat b_n+ \sum_{s} w_s  \hat b_n^\dagger  \hat b_{n+s},
 		\label{eq:BM-hami}
 	\end{gather}	
 	where $m^*$ is the effective mass at $\Gamma$, $q \in (-q_0/2,q_0/2]$ is the wave vector in the Brillouin zone, $Q_n=nq_0$ with $q_0=2\pi/Na_0$ is the $n$th reciprocal lattice site, $ \hat b_n^\dagger(q)=\sum_{Ii} e^{i(q+Q_n)R_{Ii} } \hat c^\dagger_{Ii}$. \textcolor{black}{The first term in Eq. (\ref{eq:BM-hami}) describes the dispersion of the $n$th folded subband}, $w_s$ denotes the interband scattering between the $n$th and $(n+s)$th subbands. At large $N$, the scattering strength $w_s$ dominates the $n$th subband dispersion (See Supplementary \cite{SI} S1), resulting in the $n$th flat band, As shown in Fig. \ref{fig:bandstructure}(a, b), TB and CM bands show two well-separated perfect flat bands below the dispersive bands at $N=14$ .
 
	\textit{Photonic flat bands.—}We now consider all-dielectric PhCSs consisting of 1D lattice of free-standing meta-gratings as the photonic realization of the proposed model. To realize breathing hoppings, the strip widths in different sub-lattices are assumed to be  $d_{1}=\kappa a_1$ and $d_{2}=\kappa a_2$.
	Fig. \ref{fig:bandstructure}(a, b) present the band structures numerically calculated by the guided mode expansion (GME) method \cite{RN332}. Here, the structural parameters are $a_0=190$ $nm$, $\kappa=0.6$, $h=144$ $nm$, $\delta = 0.06$. The permittivity of the dielectric is $\varepsilon_m=7.022$ ($\mathrm{TiO}_2$). As shown in the figure, a pair of flat bands are observed, which agree with TB and CM. 
	The electric and magnetic fields at $\Gamma$ shown in Fig. \ref{fig:bandstructure}(d, g) \textcolor{black}{(see Fig. S8 in Supplementary \cite{SI} for other $k$ points)} are strongly localized in the shrunken region with no (lower band) and one (higher band) node. 
	
	It is instructive to investigate the evolution of the band structure with respect to the breathing strength. Fig. \ref{fig:transision} shows the calculated photonic band structures for the TM-polarization with $N=14$ and $\delta$ varying from $0$ to $0.13$. For $\delta=0$, there are two folded dispersive bands separated by a large band gap, which are similar to the conduction- and valence-like electronic bands in semiconductors.  Away from $\delta=0$, a band gap opens at $K$ in both conduction and valence bands, resulting in a pair of isolated bands. The gap width increases further with $\delta$, as shown in Fig. \ref{fig:transision}(b). After the gap opening at $K$, the first band crossing point at $\Gamma$ opens and generates a second pair of flat bands. Further TB calculations show the rate of gap opening at $K$ is linear with $\delta$, which is faster than the rate $\propto\delta^2$ at $\Gamma$ (see Supplementary \cite{SI} S9 for derivations). Consequently, the second pair bands become flat at $\delta=0.12$ [Fig. \ref{fig:transision}(c)].
	\textcolor{black}{As $\delta$ further increases, the bandwidth of flat bands initially decreases exponentially [Fig. \ref{fig:transision} (i-j)]. When the two groups of flat bands meet at large $\delta$, the coupling between subbands leads to the band gap close and reopen sequentially at $\Gamma$ and $K$ [Fig. \ref{fig:transision}(e-h)]. This leads to the re-dispersion of flat bands. Simultaneously, new flat bands continue to emerge, indicating that a wider bandgap is favorable for generating more flat bands.}
	
\begin{figure*}
		\centering
		\includegraphics[width=\linewidth]{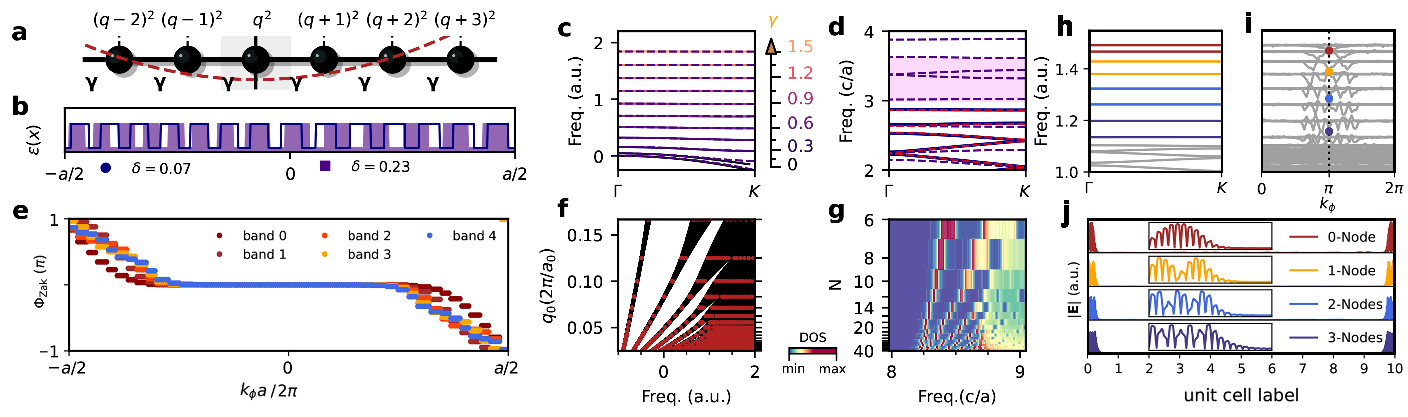}
		\caption{
			(a) The CM lattice model in momentum space. 
			(b) Permittivity distribution for $\delta=0.07$ (blue) and 0.23 (purple). 
			(c) Fitted (purple) and numerical (dashed) evolution of the zeroth flat band $E^{(0)}_q(0)$ with $\gamma$. 
			(d) Band structure calculated with the $\hat{\Theta}$ operator for $\delta=0.23$, $N=14$ (dashed purple), numerical diagonalization for $\delta=0.17$ (solid blue), and the effective model (dashed red) (see Supplementary \cite{SI} S5 for such effective model). 
			\textcolor{black}{(e) $\phi_{\text{Zak}}$ for the first $4$ bands in (h).} 
			(f) Dependence of the energy spectrum on $q_0$, where red points correspond to $q_0=2\pi/N a_0$ with even $N$. 
			(g) Dependence of DOS on frequency and $N$ for $\delta=0.03$. 
			\textcolor{black}{(h) Band structure calculated from $\hat{\Theta}$ for $N=40$, $\delta=0.15$.} 
			\textcolor{black}{(i) Energy spectrum of the finite lattice with respect to $k_{\phi}$ (see Supplementary \cite{SI} S6 for calculations' detail). }
			\textcolor{black}{(j) Topological edge states at frequencies marked in (i). Inset: Field patterns of the edge states.}
		}
		\label{fig:origin}
	\end{figure*}

	\textit{Microscopic mechanism.—}To gain quantitative insights into the underlying physics, we simplify the CM Hamiltonian by only considering the scattering between nearest-neighbor subbands [Fig. \ref{fig:origin}(a)]. For $s=1$, Eq. (\ref{eq:BM-hami}) can be recast into the iterated matrix form:
	\begin{gather}
		H_m(q)=\begin{bmatrix}
			(q-m q_0)^2&\gamma I_{m-1}&0\\
			\gamma I_{m-1}^T&H_{m-1}(q)&\gamma I_{m-1}^T\\
			0&\gamma I_{m-1}&(q+m q_0)^2
		\end{bmatrix},
		\label{eq:CM_mat}
	\end{gather}
	where $I_m=[1,0,\dots,0]$, $H_0= q^2$ and $\gamma=2m^* w_1 $ is the reduced interband scattering. The cascadingly generated flat bands can be interpreted by Eq. (\ref{eq:CM_mat}). To be specific, we note that the iterated order $m$ naturally defines two sequenced energy windows, i.e., the diagonal elements $E^{(0)}_q(m)=(q\pm m q_0)^2$. For a fixed $m$, the off-diagonal elements of $H_m(q)$ become comparable to $E^{(0)}_q(m)$ as $\gamma$ increases, suggesting band gap opens at $\Gamma$ when $\gamma$ approaches $E^{(0)}_\Gamma\propto m^2$, and at $K$ when $\gamma$ near $E^{(0)}_K\propto (\pm m + 1/2)^2$. Consequently, flat bands emerge cascadingly by continuously increasing $m$ and $\gamma$. As illustrate in Fig. \ref{fig:origin}(c), a polynomial form of $E^{(0)}_q(0)$ can also be obtained, clearly showing weak dispersion at large $\gamma$ (see Supplementary \cite{SI} S11 for detailed derivation).
	\textcolor{black}{Furthermore, as shown in Fig. \ref{fig:origin}(f), the number of flat bands also increases with $N$, which is consistent with GME calculations [Fig. \ref{fig:origin}(g)]. This dependence can be directly obtained from the dependence on $\gamma$ by renormalizing $\gamma$ in CM (see Supplementary \cite{SI} S12 for the detailed derivation)}.
	
\begin{figure}
		\centering
		\includegraphics[width=\linewidth]{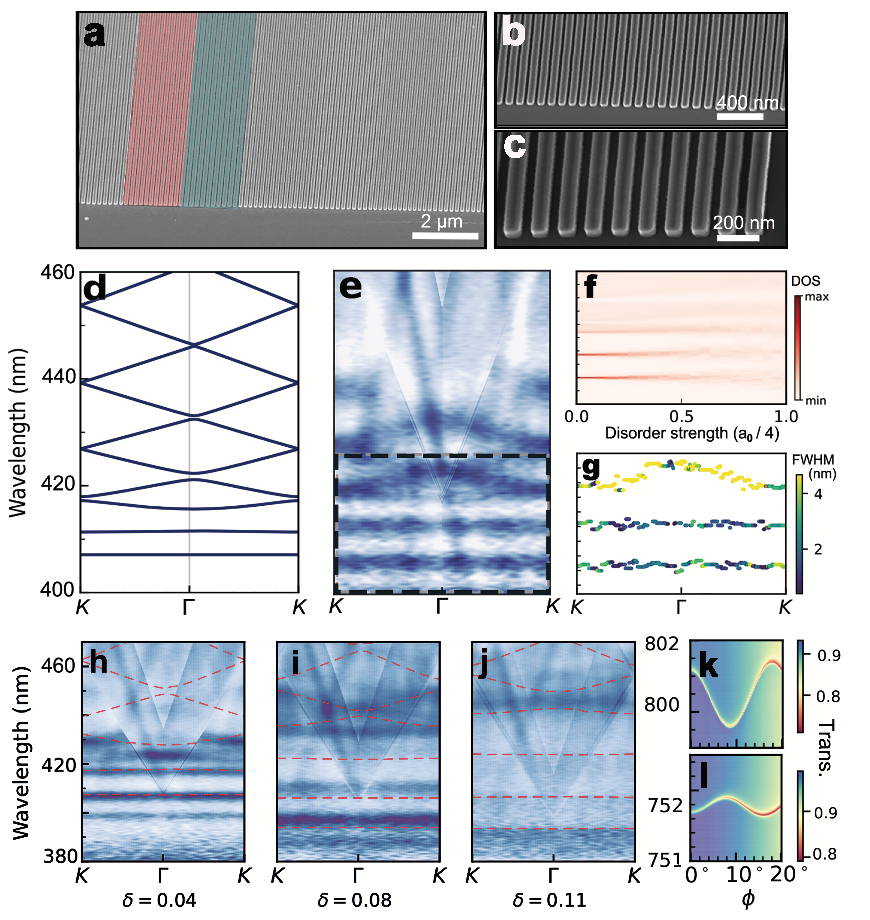}
		\caption{(a-c) SEM images of fabricated $\mathrm{TiO}_{2}$ PhCS. Simulated (d) and measured (e) photonic bands for the TM-polarization with the structure parameters $N=24, \delta=0.018, \kappa=0.6, a_0=202.95,h=144~ {nm}$ and $h_s=20~ {nm}.$  (f) Evolution of DOS of flat bands on intra-cell disorder for $100$ samples. (g) FWHM extracted from the experimentally measured spectrum (see Supplementary \cite{SI} S13),  which indicating the leakage of the eigen-modes. (f) and (g) share the same energy window marked in (e). (h-j) Measured and calculated bands (dashed) for the TM-polarization of PhCSs with $N=14$. The structure parameters are tabled in (see Supplementary \cite{SI} Tab. S2) and the dashed line are GME band structures. \textcolor{black}{(k-l) RCWA calculated high Q flat bands for $0$th and 1$st$ LLs of a free standing sample.}}
		\label{fig:expset}
	\end{figure}

	The interband scattering mechanism can be rigorously confirmed by an exact optical model (see Supplementary \cite{SI} S4-S6).  The system can be solved by considering the plane wave expansion (PWE) of the operator $\hat{\Theta} = \nabla \times (\epsilon^{-1}(\mathbf{r}) \nabla \times)$, which satisfies the eigenvalue problem $\hat{\Theta} \, \mathbf{H} = (\omega/c)^2 \mathbf{H}$. To be specific, we have the matrix elements,
	\begin{gather}
		\Theta_{ij}=h_q(g_i,g_j) \tilde\varepsilon(g_i-g_j),
		\label{eq:psudo-hamiltonlain}
	\end{gather}
	where $g_i =2i\pi/Na_0$ labels the reciprocal lattice site, and $h_q(g_i, g_j) = q^2 + q(g_i + g_j) + g_i g_j$ describes the dispersion relation of subbands ($i=j$) and interband scattering ($i\neq j$). The Fourier component of the inverse permittivity reads,
	\begin{gather}
		\tilde\varepsilon(g)=\varepsilon_m^{-1} \delta_{g,0}+ \frac{i(1-\varepsilon_m^{-1})}{g}  \Sigma_g(\delta),
		\label{eq:bands_cp}
	\end{gather}
	where $\Sigma_g(\delta)=\sum_{n=0}^{N-1}\left(e^{i\kappa g a_n}-1\right) e^{-ig R_n}$ and $a_n(\delta)=R_{n+1}-R_n$. The structural information is encoded in the interband scattering via the sum $\Sigma_g(\delta)$.
	\textcolor{black}{It is shown that the original band gap for $\delta=0$ arises from scatterings between the adjacent $N$ subbands, i.e. $\sum_n e^{ig_l R_n}=\sum_n e^{i2\pi nl/N} = \delta_{l,\mathbb{Z}N}$. For the superlattice with $\delta\neq0$, flat bands emerge from additional scatterings between \(N \pm j\) subbands with small $j\ll N$ (\(j = 1, 2, \dots\)). Additional calculations demonstrate that the re-dispersion is due to a mixing process involving scatterings at lager $j$s. Effective $w_s$ in Eq. \eqref{eq:BM-hami} can be extracted from PWE results, which is independent of momentum and consistent with the CM picture (see Supplementary \cite{SI} S4 for detailed derivation).} Additionally, Fig. \ref{fig:expset}(f) demonstrates the momentum-independent scattering mechanism resilient against weak intra-cell disorder smaller than $a_0/8$  (see Supplementary \cite{SI} S10 for details).
	
	\textit{Dual LLs and topology.---}\textcolor{black}{After establishing the microscopic origin of flat bands from the momentum-independent interband scatterings, we now exploit the global topology of the breathing superlattices. As demonstrated in Fig. \ref{fig:bandstructure}(c,d), the eigen-fields of the flat bands exhibit similar nodal structures as the field patterns of different orders of LLs for the 2D quantum dynamics of charged particles in an external magnetic field, which indicates a topological origin for these flat bands. However, our system is a 1D periodic lattice without introducing time reversal or inversion symmetry breaking, which differs from conventional Chern or valley Chern insulators. To motivate, we reconsider the CM by considering the intraband dispersion beyond the effective mass approximation,}
	\begin{gather}
		H=\sum_{n,q}t_0 \cos((q+nq_0)a_0) b_{n,q}^\dagger b_{n,q}+w_1\sum_{s=\pm1}   \hat b_{n,q}^\dagger  \hat b_{n+s,q}
		\label{eq:dualAAHmodel}
	\end{gather}
	\textcolor{black}{where $q\in[0,q_0)$. Eq. \eqref{eq:dualAAHmodel} supports an unexpected duality between our 1D system with the HH model \cite{Harper55} on the reciprocal lattice. To this aim, we adiabatically move sites inside the unit cell. With periodic boundary condition (PBC), this process brings no physical consequence except that an auxiliary pure gauge $k_\phi$ is introduced to the momentum. By substituting $q a_0\rightarrow k_m=q a_0+k_\phi\in [0,2\pi)$, a fictitious dimension can be defined by performing Fourier transform of $b_{n,k_m}^\dagger=\frac{1}{\sqrt{N_m}} \sum_me^{ik_m m} b_{m,n}^\dagger$, which yields the following lattice Hamiltonian (see Supplementary \cite{SI} S2 for derivations): }
	\begin{gather} 
		H=  \sum_{m,n} \frac{t_0}{2}  e^{in \Phi} b_{m+1,n}^\dagger b_{m,n} + w_1 b_{m,n+1}^\dagger b_{m,n} + h.c., \label{eq: HarperHofstadterModel}
	\end{gather}
	\textcolor{black}{where \( \Phi = q_0 a_0 := \Phi_0/ N \) with $\Phi_0=2\pi$, $m$ and $n$ label the reciprocal lattice points associated with the fictitious and physical dimensions, respectively. Eq. (\ref{eq: HarperHofstadterModel}) is formally equivalent to the Hamiltonian of the 2D HH model, where $\Phi$ takes the role of the magnetic flux.  In the HH model $\Phi/\Phi_0=P/Q$, where $\Phi_0$ is the flux quantum and $P, Q$ are integers \cite{Harper55}. In our 1D superlattice, we take $P=1$ with $Q = N$ representing the site number per unit cell. One of the central features of the HH model is the Hofstadter's butterfly, which is the self-similar fractal spectral energy spectrum with increasing the external magnetic field. The calculated dependence of the energy spectra on $q_0$ for $P=1$ sector clearly shows a typical Hofstadter butterfly pattern, which confirms the dual picture. In the HH model, the Hofstadter butterfly pattern reduces to LLs in the weak field limit. In our system, this picture explains the cascade LLs at large $N$, as shown in Fig. \ref{fig:origin}(h). Additionally, the LL picture fails at strong field, which corresponds to the absence of flat bands at small $N$. }
	
	\textcolor{black}{The nontrivial topology of the dual HH model with 1D edge states in the fictitious dimension provides a topological description for flat bands.  By dimensional reduction \cite{PhysRevB.78.195424}, the Chern number of the HH model reduces to the Zak phase  $\phi_{\text{Zak}}$, which is protected by inversion symmetry.
	To be specific, $\phi_{\text{Zak}}$ becomes nontrivial when two 1D edge states cross at $k_\phi=\pi$, leading to 0D edge states in the minigap [Fig. \ref{fig:origin}(h,i)]. Additionally, flat bands originating from the same conduction (or valence) bands have the same $\phi_{\text{Zak}}$, where the Wannier center locates either at the unit center ($\delta > 0$, $\phi_{\text{Zak}} = 0$) or at the boundary ($\delta < 0$, $\phi_{\text{Zak}} = \pi$), exhibiting the same node structures as LL wavefunctions [Figs. \textcolor{blue}{S2} and \textcolor{blue}{S3} \cite{SI}]. Numerical calculations for the Chern number in the 2D $(k,k_\phi)$ space  [Fig. \textcolor{blue}{S7} \cite{SI}] and $\phi_{\text{Zak}}$ at $N=40$ confirm this dual picture [Fig. \ref{fig:origin}(e)]. Consequently, topological edge states are shown in Fig. \ref{fig:origin}(j).}
	
\begin{figure}
		\centering
		\includegraphics[width=\linewidth]{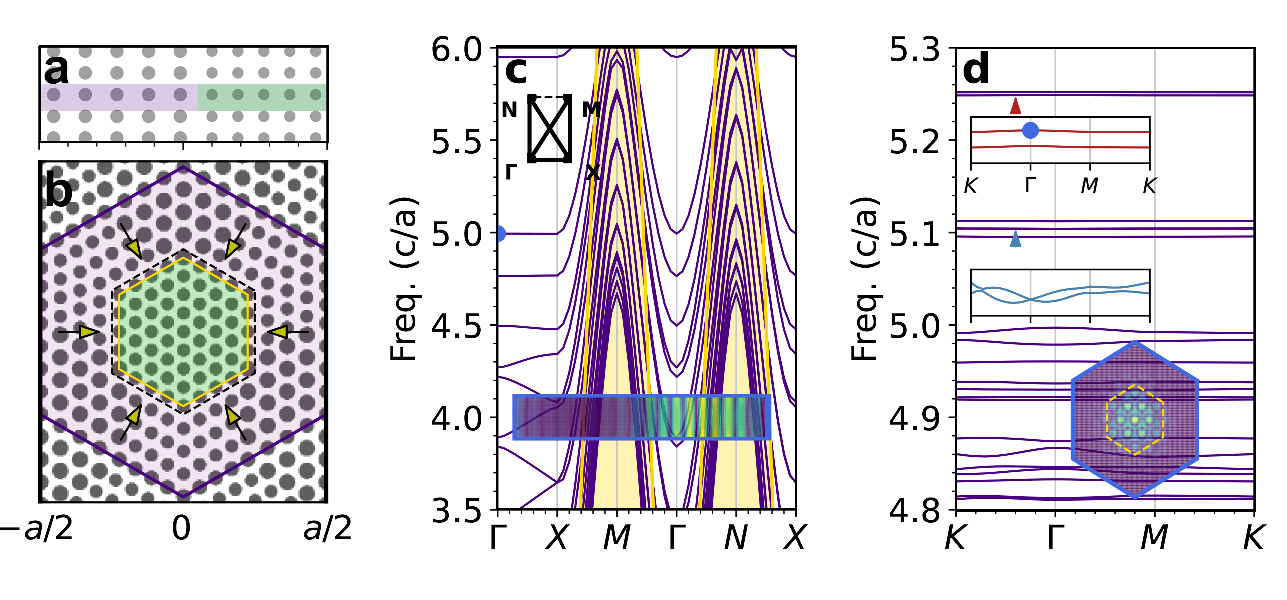}
		\caption{
			(a) Superlattice along only the $x$-directions with thickness $h=0.6a_0$.
			(b) Triangular superlattice on both directions with thickness $h=0.6a_0$. 
			(c) TM-polarized photonic band structures with partial flat bands, the shaded region represents modes below the light-line. 
			(d) Omnidirectional flat bands for a PhCS with superlattice shown in (b). Insets in (c,d) are the corresponding BZs and field distributions of $|\mathbf{H}|$ for flat bands at $\Gamma$.}
		\label{fig:2Dband}
	\end{figure}

	\textit{Experimental demonstration.—}For an experimental realization of photonic flat bands, we fabricate all-dielectric PhCSs on $\mathrm{TiO}_{2}$ membranes by using standard nanofabrication technology combining electron-beam lithography and dry etching process. Representative scanning electron microscope (SEM) images of the sample are shown in Fig. \ref{fig:expset} (a-c), where a $144$ $nm$ thick $\mathrm{TiO}_{2}$ and an intermediate ITO layer are deposited on the $\mathrm{SiO}_{2}$ substrate by using atomic layer deposition. The filling factor of the sample is $0.6$. We apply a momentum-space imaging spectroscopy system to measure the angle-resolved transmission spectrum (ARMS, Ideaoptics). In Fig. \ref{fig:expset}, the measured TM-polarized band structure shows two flat bands, which are consistent with numerical calculations in a broad frequency range.  \textcolor{black}{However, the refractive index mismatch due to the substrate and fabrication imperfection result in low Q-factors ($\sim100$) [Figs. \ref{fig:expset}(g) and also see Supplementary \cite{SI} S7.]} In addition, as shown in Fig. \ref{fig:expset}(h-j), flat bands appear cascadingly and shift towards shorter wavelength. Because the Q-factors decrease rapidly with $\delta$, the measured bands are blurred except the uppermost flat band, which remain distinct due to the weak leakage. 
	The proposed 1D breathing superlattices can be readily extended to higher dimensions. Fig. \ref{fig:2Dband}(a) presents the simplest generalization to 2D, where the breathing superlattice is along the $x$-axis and remains a simple periodic lattice along the $y$-axis. The corresponding photonic band structure for a PhCS consisting of 2D hole arrays of the proposed partial breathing superlattice milled in the	 free-standing TiO$_2$ membrane is shown in Fig. \ref{fig:2Dband} (c), where partial flat bands are observed along the $\Gamma$-$X$ direction. Similar to 1D systems, the distribution of the magnetic field shows that light is strongly localized in the shrunken region. Obviously, it is easy to realize partial flat bands in an arbitrary direction by choosing the orientation of the breathing superlattice.
	
	\textit{2D generalization.---}\textcolor{black}{To achieve omnidirectional flat bands requires structural breathing in both directions. A representative 2D breathing superlattice created from a simple triangular lattice is shown in Fig.  \ref{fig:2Dband} (b), where the expanding and compressed regions are respectively shaded with green and purple (see Supplementary \cite{SI} S14 A for construction of 2D breathing lattice). Fig. \ref{fig:2Dband} (d) presents the calculated band structures of a free-standing PhCS consisting of hole arrays milled in the membrane. It is shown that there exist two flat bands at higher frequencies spanned over the full Brillouin zone, and four weakly dispersive bands near $5.1c/a$. Interestingly, these flat bands are reminiscent of those generated from 
	$\Gamma$ valley tTMDs \cite{RN441}. Moreover, flat bands in 2D breathing superlattices emerges cascadingly when $\delta$ increases, which is similar to 1D cases (see Supplementary \cite{SI} S14 B for detailed). As shown in the inset of Fig. \ref{fig:2Dband} (d), however, these omnidirectional flat bands appear in groups due to the protection of an approximated $C_6$ symmetries, which indicates that the lattice symmetry offers an additional degree of freedom in designing flat bands in 2D. }
	
	\textit{High quality flat bands.---}\textcolor{black}{The Q-factor of optical resonances is essential for increasing the life time of quasi-particles in light-matter interactions. Meanwhile, the enhanced light-matter interactions in general flat bands systems are usually attributed to compact localization] \cite{RN661}. It is interesting to simultaneously achieve both of them in the breathing superlattices \textit{without substrate} with $\delta=0.1$ at long wavelength $\lambda\sim750 ~ \text{nm}$. As shown in Fig. \ref{fig:expset} (k, l), the transmission spectra of a free standing PhCSs is calculated through the rigorous coupled wave analysis (RCWA) method at $N=14$. Interestingly, the Q-factor of flat bands can reach $Q=1.5\times 10^5$, which is comparable to the finely tuned high quality flat bands in moir\'e lattice \cite{Nasidi:23}. The calculated high Q-factor has a similar mechanism as the quasi-bound states from band folding: when $\delta=0$, the dispersive bands are exact bound states below the light cone; While for $\delta\neq 0$, the structural breathing folds the dispersive bands above the light cone, and the inter-band scatterings introduce weak leakage into the free space, generating wide angle high-Q flat band resonances.}

	\textit{Conclusion.—}In this Letter, we have demonstrated a general strategy to construct cascading topological flat bands on demand with breathing photonic superlattices, which have an original from LLs of the 2D HH model. Our proposal can be extended to higher dimensional systems, where the anisotropy of flat bands offers another degree of freedom to tune the light-matter interaction. The proposed breathing superlattice not only provides a platform for innovative photonic applications but also applies to other physical systems such as magnons, optical lattices, acoustic and elastic waves.
	
	\acknowledgments{
		This work is supported by Chinese Academy of Sciences (CAS, No. XDB28000000, No. QYZDJ-SSWSYS001, No. XDPB22 and XDB0460000), National Natural Science
		Foundation of China (No. 61875225 and No. 12374282) and Innovation Program for Quantum Science and Technology (Grant No. 2024ZD0300104)}
	\bibliography{Photonic_flat_band.bib}
	
\end{document}


\title{Supplemental Material for Emergence of  Cascading Flat Bands in Breathing Superlattices}
	
	\affiliation{
		State Key Laboratory of Semiconductor Physics and Chip Technologies, Institute of Semiconductors, Chinese Academy of Sciences,  Beijing 100083, China
	}
	\affiliation{
		State Key Lab of Fabrication Technologies for Integrated Circuits, Institute of Microelectronics, Chinese Academy of Sciences, Beijing 100029, China
	}
	
	\affiliation{
		Center for Quantum Matter, Zhejiang University, Hangzhou 310027, China
	}
	\affiliation{
		College of Materials Science and Opto-electronic Technology, University of Chinese Academy of Sciences, Beijing 100049, China
	}

	\author{Moru Song\orcidlink{0009-0003-4842-6959}}
	\affiliation{
		State Key Laboratory of Semiconductor Physics and Chip Technologies, Institute of Semiconductors, Chinese Academy of Sciences, Beijing 100083, China
	}
	\affiliation{
		College of Materials Science and Opto-electronic Technology, University of Chinese Academy of Sciences, Beijing 100049, China
	}
	\author{Jinyu Hu}
	\affiliation{
		State Key Lab of Fabrication Technologies for Integrated Circuits, Institute of Microelectronics, Chinese Academy of Sciences, Beijing 100029, China
	}
	\author{Lina Shi}
	\email{shilina@ime.ac.cn}
	\affiliation{
		State Key Lab of Fabrication Technologies for Integrated Circuits, Institute of Microelectronics, Chinese Academy of Sciences, Beijing 100029, China
	}
	
	\author{Yongliang Zhang\orcidlink{0000-0001-9419-6881}}
	\email{ylzhang@semi.ac.cn}
	\affiliation{
		State Key Laboratory of Semiconductor Physics and Chip Technologies, Institute of Semiconductors, Chinese Academy of Sciences, Beijing 100083, China
	}
	\author{Kai Chang\orcidlink{0000-0002-4609-8061}}
	\email{kchang@zju.edu.cn}
	\affiliation{
		Center for Quantum Matter, Zhejiang University, Hangzhou 310027, China
	}
	
	\date{\today}
	\maketitle
	\tableofcontents

\section{Derivation of the continuum model }
Here, we first rewrite the tight-binding (TB) Hamiltonian  Eq. (\ref{main-TB_hamiltonian}) by using  $\hat c^\dagger_{Ii}\hat  c_{Jj}=\ket{\phi_{I,i}}\bra{\phi_{J,j}}$,
\begin{gather}
	\hat H= t\sum_{IJ,ij} e^{-\xi(L_I-L_J+\tau_i-\tau_j)}\ket{\phi_{I,i}}\bra{\phi_{J,j}},
	\label{eq:TB2}
\end{gather}
where $\phi_{I,i}=\phi(r-L_I-\tau_i)$ is the localized Wannier basis, $R_{Ii}$ is replaced by $L_I+\tau_i$ with $L_I$  the position of the $I$th unit cell and $\tau_i$  the position of $i$th site ($0\le i,j<N$).   

Then, we use the truncated plane wave method \cite{RN596} to drive the full continuum model (CM) Hamiltonian. The plane wave basis are given by, 
\begin{gather}
	\ket{\psi_n(k)}=\frac1{\sqrt{N_m N}}\sum_{I,i} e^{i(k+Q_n)(L_I+\tau_i)}\ket{\phi_{Ii}},
	\label{eq:CMbasis}
\end{gather}
where $Q_n$ is the $n$th reciprocal lattice in the momentum space, $k\in(-\pi/Na_0,\pi/Na_0]$, $N_m$ is the number of unit cells and $N$ is the number of sites per unit cell.  By using the complete relationship of the plane wave basis, we can express Eq. (\ref{eq:TB2}) as,
\begin{gather}
	\hat H=\sum_{nm} H_{nm}\ket{\psi_m(k)}\bra{\psi_n(k)},
\end{gather}
where $H_{nm}=\bra{\psi_n(k)}\hat H\ket{\psi_m(k)}$ is the matrix element of the CM Hamiltonian,
\begin{gather}
	H_{nm}=\sum_{ij} P_{ni}^*\left( t e^{-\xi|\tau_{ij}|} e^{-i k\tau_{ij}}\right)P_{mj}.
	\label{eq:TBtoCM_numerically}
\end{gather}	
Here, $\tau_{ij}=L_I-L_J+\tau_i-\tau_j$ and the projection matrix is given by
\begin{gather}
	P_{mj}= \frac{\exp(iQ_m \tau_i)}{\sqrt{N}}. 
	\label{eq:projection_marix}
\end{gather}	
\begin{figure}
	\centering
	\includegraphics[width=\linewidth]{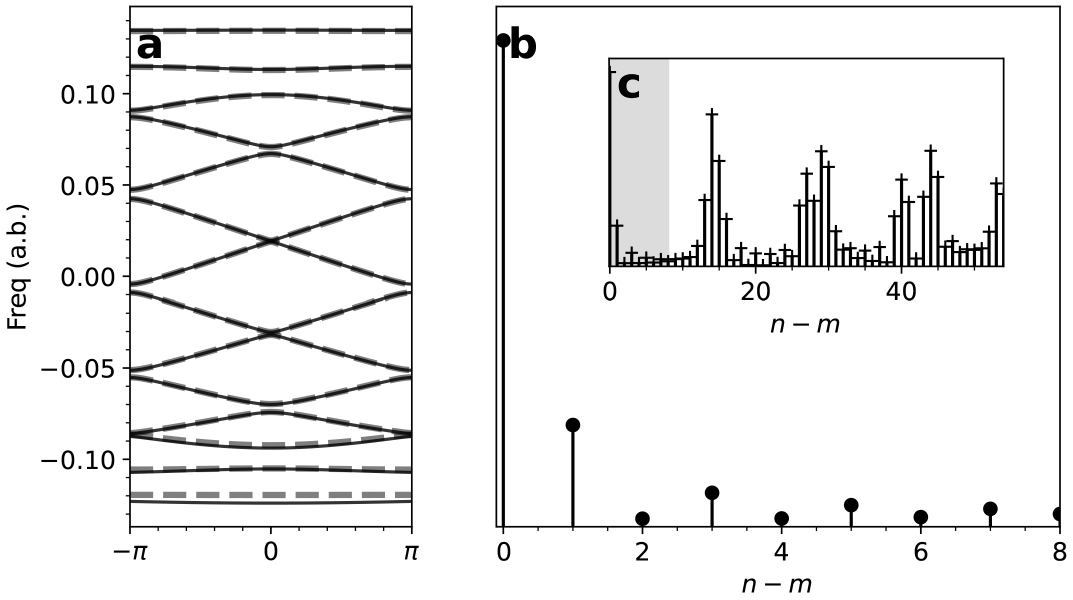}
	\caption{(a) The band structure calculated by TB (dashed) and full-CM (soild). The number of plane waves is $\sim 140$ calculated in $N=14,\delta=0.05$ case. (b,c) Scattering strength among bands defined by the matrix element of Hamiltonian Eq. \ref{eq:TBtoCM_numerically}: $S(n-m)=\sum_m |H_{m,n-m}|$. }
	\label{fig:s1tbtocm}
\end{figure}

Note that Eq. (\ref{eq:TBtoCM_numerically}) can be  numerically calculated and gives all  matrix elements of the full CM Hamiltonian. As shown in Fig. \ref{fig:s1tbtocm}(a), the numerical result shows good agreement with TB. 
Furthermore, Eq. (\ref{main-eq:BM-hami}) can be analytically obtained by considering,
\begin{gather}
	H_{nm}=\frac{t}{N}\sum_{i} e^{-i(Q_n-Q_m)\tau_i} \sum_{\tau_{ij}} e^{-i(k+Q_m)\tau_{ij}} e^{-\xi|\tau_{ij}|}.
	\label{eq:TB_to_CM_ana1}
\end{gather}

For the superlattice with different lattice spacings $a_{1,2}=a_0(1\pm\delta)$, and the nearest-neighbor hopping i.e. $2\xi a_0\gg1$, Eq. (\ref{eq:TB_to_CM_ana1}) can be expanded as follows,

\begin{align}
	H_{nm}= &\frac t N \sum_{s=1}^{n'} 2e^{-i(Q_n-Q_m)sa_0(1+\delta)} e^{-\xi a_0(1+\delta)}\nonumber\\
	&\times\cos\left[(k+Q_m)(1+\delta)a_0\right] \nonumber\\
	+\sum_{s'=n'}^{N} &2e^{-i(Q_n-Q_m)[n'a_0(1+\delta)+(s'-n')(1-\delta)a_0]} e^{-\xi a_0(1+\delta)}\nonumber\\
	&\times\cos\left[(k+Q_m)(1-\delta)a_0\right],
	\label{eq:TB_to_CM_ana2}
\end{align}
where we have defined $2n'=N$. In the limit $\delta=0$, from Eqs. (\ref{eq:TB_to_CM_ana1}) and (\ref{eq:TB_to_CM_ana2}), we can obtain 
\begin{align}
	H_{nm}^{(0)}&=t e^{-\xi a_0}\sum_{s} e^{-i(Q_n-Q_m)sa_0} \cos((k+Q_m)a_0)\nonumber\\
	&=t e^{-\xi|a_0|} \cos((k+Q_m)a_0)\delta_{Q_n,Q_m}.
	\label{eq:TB_to_CM_ana3}
\end{align}
Here, the identity $\sum_{s} e^{-i(Q_n-Q_m)sa_0}=\delta_{Q_n,Q_m}$ is used. Unlike the TBG system, where flat bands arise from the K-valley \cite{RN312}, here we only focus the physics near the $\Gamma$ valley. It is convenient to write:
\begin{gather}
	k=q+\Gamma=q.
\end{gather}
By expanding Eq. (\ref{eq:TB_to_CM_ana3}) to the second order, we obtain,
\begin{gather}
	H_{nm}^{(0)}\approx t e^{-\xi|a_0|}\left(1+\frac{(q+Q_m)^2}{2/a_0^2}\right):=\mu+\frac{(q+Q_m)^2}{2 m^*},
\end{gather}
which is the first term of Eq. (\ref{main-eq:BM-hami}). When $\delta\neq0$, we define the second term as the interband scattering,
\begin{gather}
	U_{mn}=H_{mn}-H_{nm}^{(0)}.
	\label{eq:substraction}
\end{gather}
Eq. (\ref{eq:substraction}) is difficult to  obtain directly. Here we find the approximated expression as
\begin{align}
	U_{mn}(k) &=\frac {tf_{m}(k)} N \sum_{s=1}^{n'}\left(e^{\left(-i(Q_n-Q_m)s-\xi\right)\delta a_0} -1\right)\nonumber\\
	&\times e^{-i(Q_n-Q_m)sa_0} 	
	\nonumber\\
	&+e^{-i(Q_n-Q_m)n'a_0}\left(e^{\left(-i(Q_n-Q_m)(n'-s)+\xi\right)\delta a_0}-1\right)\nonumber\\
	&\times e^{-i(Q_n-Q_m)sa_0}.
	\label{eq:TB_to_CM_ana4}
\end{align}
Since $\delta$ is small and $\cos((k+Q_m)(1\pm\delta)a_0)\approx\cos((k+Q_m)a_0)$, we have:
\begin{align}
	f_{m}(k)&=2e^{-\xi a_0} \cos((k+Q_m)a_0)\\\nonumber&=2e^{-\xi a_0} \cos( 2\pi m/N +k a_0).
\end{align}
It is important to note that $f_m(k)$ becomes independent from $k a_0\in(-\pi/N,\pi/N]$ when $m\neq \mathbb{Z} N$ with large $N$. Considering $Q_m=\frac{2\pi m}{N a_0}=m q_0$, we can express Eq. (\ref{eq:TB_to_CM_ana4}) as
\begin{gather}
	U_{mn}=\frac {tf_{m}(k)} N \sum_{s=1}^{n'} e^{-i 2\pi h s/N} g_{s,h}(\delta),
	\label{eq:TB_to_CM_ana5}
\end{gather}
where $h=n-m$ and $ g_{s,h}=(e^{-i\delta (2\pi hs/N-\xi a_0) }-1)+$ 

$ e^{-i\pi h}(e^{\delta \left(i 2\pi hs/N-i h\pi+ \xi a_0  \right)}-1).$ Since $\delta$ is small, $g_{s,h}$ is slowly varying with $s,h$. If $g_{s,h}$ is a  constant, $U_{mn}$ reaches maximum when $h=\mathbb{Z}N$, and zero in other cases for the phase matching condition. Therefore, the dependence of $g_{s,h}$ on $h,s$  breaks this condition slightly, making $h=\mathbb{Z}N+ j$ non-vanishing (with $j=\pm1,\pm2 ...$). This physical picture can also be found in Fig.  \ref{fig:s1tbtocm}(c). As a result, it is convenient to write Eq. (\ref{eq:TB_to_CM_ana5}) into
\begin{gather}
	U_{mn}\simeq \sum_{|j|<\mathrm{cut}} w_{j} \delta_{m-n,\mathbb{Z}N+ j },
	\label{eq:CM_interband}
\end{gather}
where $w_{j}\propto t$ is independent from $k$ and only depends on $N$. Since $\cos((k+Q_m)a_0)=\cos((k+Q_m)a_0+2\pi\mathbb{Z}N)=\cos((k+Q_{m+\mathbb{Z}N})a_0)$, we can substitute $Q_m\rightarrow Q_{m+\mathbb{Z}N}$ without changing the Hamiltonian.
Therefore, Eq. (\ref{eq:CM_interband}) can be recast into
\begin{gather}
	U'_{mn}= \sum_{|s|<\mathrm{cut}} w_{s} \delta_{m-n,s },
	\label{eq:CM_interband2}
\end{gather}
which is just the inter-band scattering term in Eq. (\ref{main-eq:BM-hami}). Additionally, as shown in Fig. \ref{main-fig:bandstructure}, Eq. (\ref{eq:CM_interband2}) can be directly fitted with the optical simulation. Note that the present method is analogous to the rigorous treatment of Maxwell equation in Eq. (\ref{eq:bands_cp}), which we will show in Section S4.

Also note that the projection matrix $P_{mj}$ Eq. (\ref{eq:projection_marix}) has a dimension of $N_Q\times N $, therefore the Hamiltonian $H_{nm}$ has a dimension of $N_Q\times N_Q$ and usually $N_Q>N$. Additional null state with zero energy will be introduced which should be removed finally. In addition, energy scale also changes, and this can be fixed by multiply $N_Q/N$ to $H_{nm}$.

\textcolor{black}{\section{Duality with Harper-Hofstadter model}
We begin with the standard Harper-Hofstadter (HH) model Eq. (\ref{main-eq: HarperHofstadterModel}) in the main text, 
\begin{gather} 
	H=t_0 \sum_{m,n} e^{in \Phi} b_{m+1,n}^\dagger b_{m,n}+ w_1 b_{m,n+1}^\dagger b_{m,n}+h.c., \nonumber
\end{gather} 
where $\Phi$ is the magnetic flux, and $t_0$ and $w_1$ are the hopping energies along the $m$ and $n$ directions, respectively. We perform a Fourier transform with respect to $m$ while leaving $n$ unchanged: 
\begin{gather} 
	b_{m,n}=\frac{1}{\sqrt{N_m}}\sum_{k_m} e^{-ik_m m} b_{k_m,n} \label{eq:DualBasis} 
\end{gather} 
By inserting Eq. (\ref{eq:DualBasis}) into Eq. (\ref{main-eq: HarperHofstadterModel}), we obtain 
\begin{align} 
	H&=2t_0 \sum_{n,k_m} \cos(k_m+n\Phi) b_{k_m,n}^\dagger b_{k_m,n}\nonumber\\ 
	&+ w_1 b_{k_m,n+1}^\dagger b_{k_m,n}+h.c., 
	\label{eq:AAHmodel} 
\end{align} where $k_m \in [0,2\pi)$. If we let $\Phi=q_0 a_0$ and $t_0= t e^{-\xi |a_0|}/2$, and since $Q_n=nq_0$, Eq. (\ref{eq:AAHmodel}) formally return to the CM model derived in the last section: 
\begin{align} 
	H&=t \sum_{n,k} e^{-\xi|a_0|} \cos\left[(k+Q_n)a_0\right] b_{k,n}^\dagger b_{k,n} \nonumber\\ 
	&+ w_1 b_{k,n+1}^\dagger b_{k,n}+h.c., 
\end{align} 
where we translate the matrix representation Eq. (\ref{eq:TB_to_CM_ana3}) and Eq. (\ref{eq:CM_interband2}) to a bosonic operator representation, $\delta_{Q_n,Q_m}\rightarrow \sum_n b_{k,n}^\dagger b_{k,n}$ and $\delta_{m-n,s }\rightarrow \sum_n b_{k,n+s}^\dagger b_{k,n}$ and limit $|s|=1$.}

\begin{figure}
	\centering
	\includegraphics[width=\linewidth]{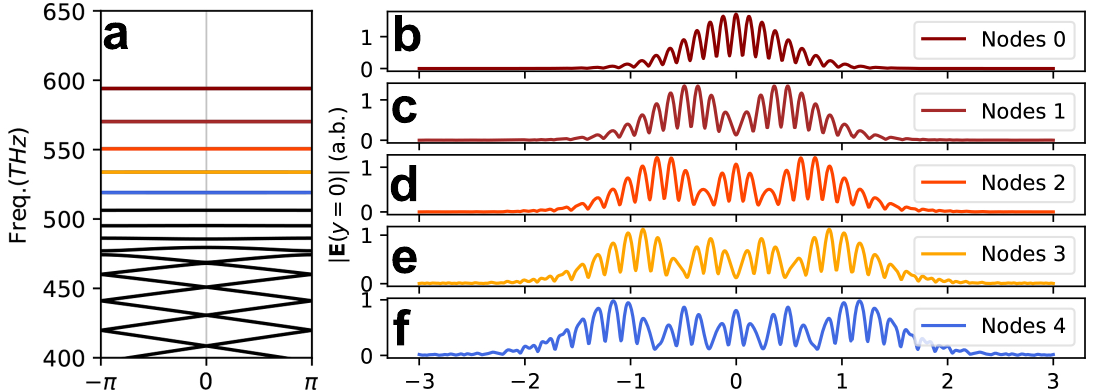}
	\caption{\textcolor{black}{(a) Band structure at $N=40$, $\delta=0.1$ for TE polarization calculated by GME (b-f) Multi band Wannier center distribution (at $y=0$ slice) in unit cell for lower Bloch bands in color, Zak phases are all $0$, the x-axis is in unit of $\mu m$.}}
	\label{fig:largen}
\end{figure}

\textcolor{black}{However, this is not the case since $k a_0 \neq k_m$, where $k a_0 \in [0,2\pi/N)$. Therefore, we should introduce an additional phase factor into this term as $k_m=k a_0+k_\phi$. Obviously, $k_\phi$ is a pure gauge that has no physical consequence. Then we have: 
\begin{align} 
	H&=t \sum_{n,k_m} e^{-\xi|a_0|} \cos(k_m+nq_0a_0) b_{k_m,n}^\dagger b_{k_m,n} \nonumber\\ 
	&\quad + w_1 b_{k_m,n+1}^\dagger b_{k_m,n}+h.c., 
	\label{eq:DualCM} 
\end{align} 
which is exactly dual to Eq. (\ref{eq:AAHmodel}). In practice, the introduction of $k_\phi$ can also be understood as shifting the unit cell in real space. Theoretically, it means that when we perform the Fourier transform on Eq. (\ref{eq:TB2}) using Eq. (\ref{eq:CMbasis}), we have an additional phase factor: 
\begin{gather} 
	\ket{\psi_n(k)}'=e^{i\theta_n}\ket{\psi_n(k)}, \bra{\psi_m(k)}'=e^{-i\theta_m}\bra{\psi_m(k)}
	\nonumber
\end{gather} 
This is consistent with Eq. (\ref{eq:DualCM}) we obtained by defining $k_\phi=\theta_{m}-\theta_{n}$. As additional evidence of the above dual picture, we have computed the Zak phase along different $k_\phi$ and the distribution of the Berry curvature in $(k,k_\phi)$ space in Fig. \ref{fig:topophasen40}, where the Chern number is $+1$ for the $N=40$ upper flat bands in Fig. \ref{fig:largen}(a). When we calculate these topological quantities as optical problems, the way to involve $k_\phi$ is by periodically shifting strips in the unit cell. }

\begin{figure}
	\centering
	\includegraphics[width=\linewidth]{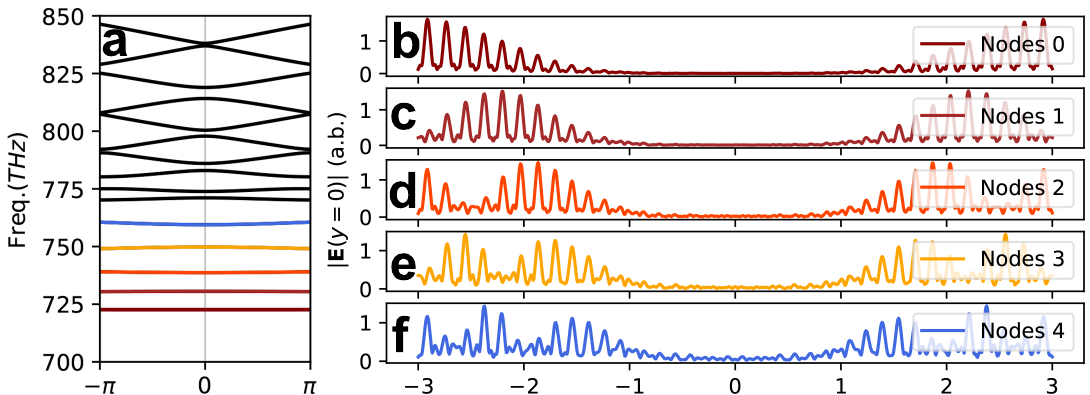}
	\caption{\textcolor{black}{Similar results at different frequency of Fig. \ref{fig:largen2}, with Zak phases are all $\pi$.}}
	\label{fig:largen2}
\end{figure}

\section{Zak phase and edge states}
\begin{figure}[htbp]
	\centering
	\includegraphics[width=\linewidth]{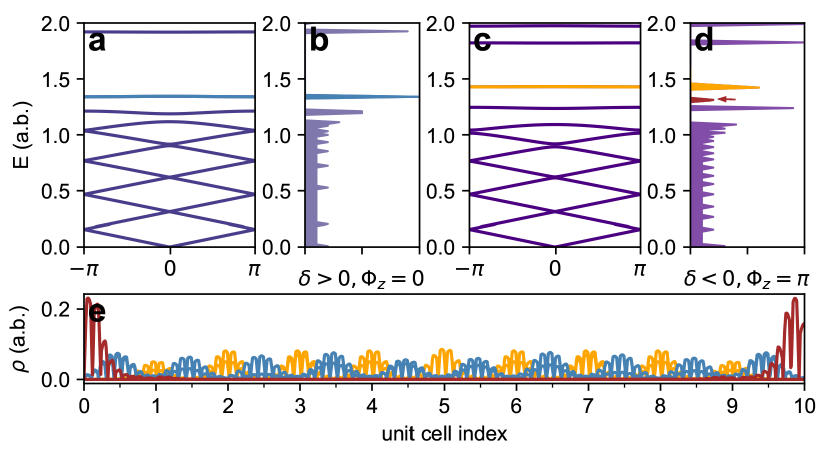}
	\caption{(a,c) Photonic band structures and (b,d) density of states of periodic and open breathing superlattices with $N=10$. Compared with the topological trivial case of $\delta>0$ (a,b), $\delta<0$ (c,d) has non-trivial topology with edge states found in the breathing lattice  with finite length (red in (d)). (e) The corresponding eigen-field distributions for the bulk and edge states. }
	\label{fig:topozakpwe}
\end{figure}
The bands in our system are characterized by a $\mathbb{Z}_2$ Zak phase depending on the sign of shift parameter $\delta$. The Zak phase for the $n$th band is defined as,
\begin{gather}
	\Phi_z=\int_{k\in \mathrm{Bz}} d k \langle{u_{kn}}| \hat x | {u_{kn}}\rangle =\bar{x}_n
\end{gather}
where $\hat x =i \partial_k$ is the position operator and $u_{kn}(x)=\psi_{nk}(x) e^{-ikx}$ with $\psi_{nk}(x)$ are Bloch wave function, this can be obtain by either eigenvectors of Hamiltonian or $\Theta$ operator (Equation of motion). Physically, $\bar{x}_n$ is the Wannier center of the $n$th band.

As illustrate in the Fig. \ref{fig:topozakpwe}, the band highlighted in blue has a trivial Zak phase $\Phi_z=0$ for $\delta>0$, while the orange one has a nontrivial Zak phase $\Phi_z=\pi$ for $\delta<0$. Since the physical meaning of Zak phase is the charge polarization, we plot the modes distribution in Fig. \ref{fig:topozakpwe}(e). The trivial Zak phase means that the Wannier centers are located at the center of the unit cell (blue), while the non-trivial Zak phase means that the Wannier centers are located at the edge of the unit cell (orange). As a further demonstration of the nontrivial topology, we have also plotted the edge states localized at the boundaries of a finite lattice for the mode marked with red in Fig. \ref{fig:topozakpwe}(d, e).  

See next section for how to obtain eigenvector  $u_{kn}$ of $\Theta$ operator in a real space representation.

\section{$\Theta$-operator in 1D}
\begin{figure*}
	\centering
	\includegraphics[width=\linewidth]{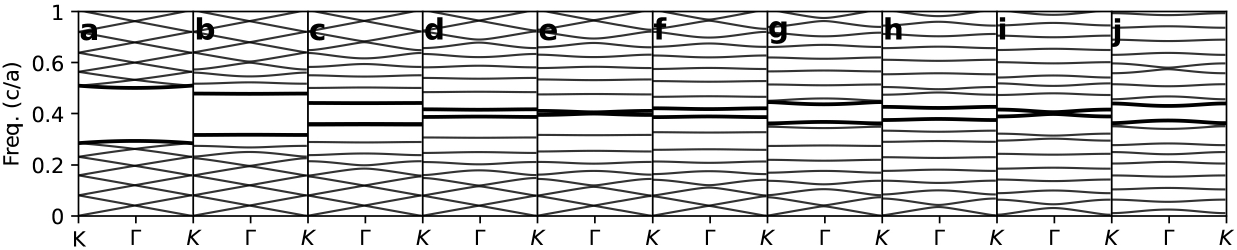}
	\caption{Evolution of the band structure from the $\Theta$ operator.}
	\label{fig:s1transision}
\end{figure*}

CM only describes the formation of flat bands. To understand the band evolution completely, We derive the ${\Theta}$ operator \cite{RN590} through the plane wave expansion (PWE) for the Maxwell equation. The  ${\Theta}$ operator defines the eigenvalue problem:
\begin{gather}
	\hat \Theta(\mathbf{r})\mathbf{H(\mathbf{r})}=\frac{\omega^2}{c^2}\mathbf{H(r)}
\end{gather}
where $\hat \Theta(\mathbf{r})=\nabla\times\varepsilon(\mathbf{r})^{-1}\nabla\times$ and $\mathbf{H(\mathbf{r},t)}=\mathbf{H(\mathbf{r})}e^{-i\omega t} $ is the stationary solution of the magnetic field. To make the derivation clearer, here we use abstract indices and set $c=1$. Then, the equation reads:
\begin{gather}
	\epsilon^{\mu \nu \rho}\epsilon_{\tau\kappa\nu}\nabla_\mu
	[\varepsilon(r)^{-1}\nabla^\tau H(r)^k]=\omega^2 H(r)^\rho
\end{gather}
with $\epsilon^{\mu \nu \rho}$ is the anti-symmetric tensor.

For the periodic system,  the Bloch theorem enables the expansion of the magnetic field:
\begin{gather}
	H^\kappa(r)=\sum_{q \in 1BZ}u_q^\kappa(r) e^{iq\cdot r}
	\label{eq:Bloch_function}
\end{gather}
Inserting Eq. (\ref{eq:Bloch_function}) into the Maxwell equation, we have:
\begin{gather}
	\epsilon^{\mu \nu \rho}\epsilon_{\tau\kappa\nu}
	(\nabla_\mu\varepsilon^{-1}(r)\cdot[\nabla^\tau+iq^\tau]+\varepsilon^{-1}(r)\times\nonumber\\
	\left[\nabla_\mu\nabla^\tau+iq^\tau\nabla_\mu+iq_\mu\nabla^\tau+q_\mu q^\tau\right] )u_q^\kappa(r)=\omega^2 u_q^\rho(r)
	\label{eq:Expanded_Maxwell}
\end{gather}
Then we use the plane wave basis to expand the vector field $u_q^\kappa$ and the inverse dielectric constant
\begin{align}
	u_q^\kappa(r)&=\sum_{i}\tilde u^\kappa(g_i+q) e^{ig_i\cdot r},\\
	\varepsilon^{-1}(r)&=\sum_i\tilde\varepsilon(g_i) e^{ig_i\cdot r}.
\end{align}
Inserting the above equations into Eq. (\ref{eq:Expanded_Maxwell}) leads to:
\begin{gather}
	-\sum_i\epsilon^{\mu \nu \rho}\epsilon_{\tau\kappa\nu} 
	\tilde\varepsilon(g_j-g_i) \\\nonumber
	\times\left[g_{j\mu} g_i^{~\tau}+q^\tau g_{j\mu}+q_\mu g_i^{~\tau}+q^\tau q_\mu\right] \tilde u^\kappa(g_i) =\omega^2\tilde u^\rho(g_j+q),
\end{gather}
where $g_i$ and $g_j$ are reciprocal lattice vectors of our system. We apply a finite cutoff to them, thus providing a finite energy window. Here, we label these reciprocal lattice vectors as $g_i$ using indices ${i,j,k}$, and then we can obtain the matrix form of our system:
$
[ \Theta_\kappa^{~\rho}]^{i}_{~~j}(q)\tilde u(q)^{\kappa j}=\omega(q)^2 \tilde u(q)^{\rho i},
$
where we have rewrite the Bloch function. Here,
$
\tilde u^{\kappa i}(q):=\tilde u^\kappa(g_i+q)
$
and the $\Theta$ matrix is defined by:
\begin{gather}
	[\Theta^\rho_{~~\kappa}]_{ij}(q)=-
	\epsilon^{\mu \nu \rho}\epsilon_{\tau\kappa\nu} 
	\tilde\varepsilon(g_i-g_j)\\\nonumber
	\times\left[g_{i\mu} {g_j}^\tau+q^\tau g_{i\mu}+q^\tau g_{j\mu}+q^\tau q_\mu\right].
\end{gather}
For the 1D case, $g_y$ and $g_z$ are zero, so both $\mu$ and $\tau$ are $x$. For the anti-symmetric tensor, we use $\nu=y$ and $\rho=\kappa=z$ for the TE mode, or $\nu=z$ and $\rho=\kappa=y$ for the TM mode. The explicit formulae for both polarizations are identical, and the $\Theta$ operator is given by:
\begin{gather}
	\Theta_{ij}(q)=\left[\left(q+\frac{g_{i}+ g_{j}}{2}\right)^2+\frac{3g_{i}g_{j}}{4}\right]\tilde\varepsilon(g_i-g_j).
	\label{eq:1Dthetamat}
\end{gather}
In more compact form, $\hat \Theta$ of the 1D breathing superlattice is given by
\begin{gather}
	\hat\Theta=\sum_{ij} h_q(g_i,g_j) \tilde\varepsilon(g_i-g_j) \hat b_i^\dagger \hat b_j,
	\label{eq:psudo-hamiltonlain}
\end{gather}
where $g_i =2i\pi/Na_0$ labels the reciprocal lattice site, and $h_q(g_i, g_j) = q^2 + q(g_i + g_j) + g_i g_j$ describes the dispersion relation of subbands ($i=j$) and interband scattering ($i\neq j$).
The dielectric distribution in Fig. \ref{main-fig:origin}(b) can be written as: $\varepsilon(x)=\sum_i \varepsilon_{m} Rect(x-R_i,\kappa a_i) +\varepsilon_{0}$
with the rectangular function
\begin{gather}
	Rect(x,a)=
	\begin{cases}
		1 \quad 0<x<a\\
		0 \quad others
	\end{cases}.
	\label{eq:rect}
\end{gather}
The Fourier transformation of the rectangular function Eq. (\ref{eq:rect}) is $\frac{i}{g} (e^{ia g}-1)$. The inverse of this rectangular permittivity function is given by  
\begin{gather}
	\varepsilon(x)^{-1} = c+b\sum_n\mathrm{Rect}\left((x-R_i),\kappa a_i\right),
\end{gather}
where $c=\varepsilon_0^{-1}$, $b=\varepsilon_m^{-1}-\varepsilon_0^{-1}<0 $ and $a_i = R_{i+1}-R_{i}$. From Eq. (\ref{eq:1Dthetamat}), we have $\tilde\varepsilon(g)=\mathcal{F}[\varepsilon(x)^{-1}](g)$.
\begin{figure}
	\centering
	\includegraphics[width=\linewidth]{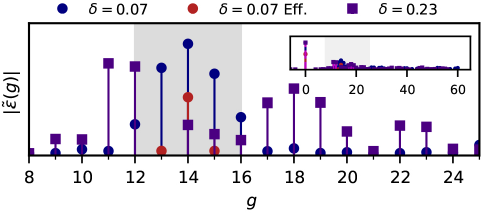}
	\caption{$\tilde\varepsilon(g)$ distribution for $\delta=0.07$ (From $\Theta$ Operator in blue, effective model red) and $\delta=0.23$ (purple), inset: extended scale.}
	\label{fig:originSM}
\end{figure}
Then, we obtain the explicit expression of $\tilde\varepsilon(g)$,
\begin{gather}
	\tilde\varepsilon(g)=\varepsilon_m^{-1} \delta_{g,0}+ \frac{i(1-\varepsilon_m^{-1})}{g}  \Sigma(g),
	\label{eq:bands_cp}
\end{gather}
where $\Sigma(g)=\sum_{n=0}^{N-1}\left(e^{i\kappa g a_n}-1\right) e^{-ig R_n}$, $a_n=R_{n+1}-R_n$ is the site spacing. The structural information is encoded in the interband scattering via the sum $\Sigma(g)$. For $\delta=0$, the formation of band gaps comes from scatterings between the $i$th and $(i+tN)$th bands with $t\in \mathbb{Z}$ because of the interference condition (IFC) $\sum_n e^{ig_l R_n}=\sum_n e^{i2\pi nl/N} = \delta_{l,tN}$. For $\delta\neq 0$, there are additional scatterings between the $i$th and $i\pm(tN\pm j)$th bands $(j=1,2,3,\cdots)$, which leads to the  formation and evolution of flat bands. This can be found in Fig. \ref{main-fig:origin}(d), whose $\tilde\varepsilon(g)$ are shown in Fig. \ref{fig:originSM}. When $\delta=0.07$, IFC  breaks slightly, and interband scatterings appear at $j=1,2$, leading to two flat bands. However, for $\delta=0.23$, the strongly broken IFC leads to multi-band scatterings involving large $j$, resulting in the re-dispersion of flat bands. Interestingly, it is sufficient to only consider $i\pm(N\pm 1)$th bands scattering to generate flat bands, as shown in the minimal effective model in Fig. \ref{main-fig:origin}(d) . Moreover, from the expression of $h_q(g_i, g_j)$, the dispersion effect is negligible in the scattering between the $i$th and $i\pm(N\pm1)$th bands when $g \gg q$, which validates the assumption of momentum-independent interband scattering at large $N$. 

Additionally, this results can capture all the evolution properties of flat bands, as illustrated in Fig. \ref{fig:s1transision}.
\section{A Minimal effective model}

As discussed in the main text, the formation of flat bands is attributed to the momentum-independent scattering  due to the structural perturbation in breathing superlattices. Eqs. (\ref{eq:psudo-hamiltonlain}) and (\ref{eq:bands_cp}) provide explicit expressions for the scattering strength between the $i$th and $j$th flat bands as $h_q(g_i,g_j)$. Here, we propose a minimal effective model that captures the emergence of flat bands by selectively including only the scattering processes between the $i$th and the $i\pm(N\pm1)$th sub-bands. For the 1D breathing superlattice, the model is obtained as:
\begin{gather}
	\tilde\varepsilon(g)=c\delta_{g,0}+ \sum_{t\in \mathbb{Z}}\sum_{s=0}^{N-1} d_{t,s} \delta_{g,(tN+s)g_0},
	\label{eq:mini-model1}
\end{gather}
where $d_{t,u}$ are parameters depending on $\delta$. Eq. (\ref{eq:mini-model1}) is equivalent to Eq. (\ref{eq:bands_cp}), and $d_{t,u}$ can be regarded as the scattering strength directly. Inserting Eq. (\ref{eq:mini-model1}) into Eq. (\ref{eq:1Dthetamat}) yields,
\begin{gather}
	\hat\Theta(q)=\sum_{i} c(q+g_i)^2 \hat b_i^\dagger \hat b_i  \\\nonumber
	+\sum_{t,s} d_{t,u}h_q(g_i,g_i+(tN+s)g_0) \hat b_i^\dagger \hat b_{i+tN+s}.
	\label{eq:mini-model2}
\end{gather}

For simplicity, we only consider the scattering among the $i$th, $i\pm N$th, and $i\pm (N\pm1)$th  bands, and then the effective minimal model is given by:
\begin{gather}
	\hat\Theta(q)=\sum_{i=0} c (q+g_{i})^2 \hat b_{i}^\dagger \hat b_{i} +\sum _{s=-1,0,1}w_{s}(q) \hat b_{i}^\dagger \hat b_{i\pm (N+ s)},
	\label{eq:mini-model3}
\end{gather}
where $w_0=d_{1,0}h_q(g_i,g_i+N)g_0)$, $w_{-1}=d_{0,N-1}h_q(g_i,g_i+(N-1)g_0)$ and $w_{1}=d_{1,1}h_q(g_i,g_i+(N+1)g_0)$, respectively. Eq. (\ref{eq:mini-model3}) has the similar form with CM. Obviously, we can find that $h_q(g_i,g_i+Ng_0)=q^2+q(2g_i+Ng_0) + g_i(g_i+Ng_0))$ is almost independent from $q$ when $N$ is large, which confirms the weak momentum dependence of superlattice scattering. Fig. \ref{main-fig:origin}(d) presents the results for the effective model, while Fig. \ref{main-fig:origin}(e) shows the parameters $c = 0.28$, $d_{1,0} = 0.176$, $d_{0,N-1} = d_{1,1} = 0.014$, which were fitted with the direct diagonalization of Eq. (\ref{eq:psudo-hamiltonlain}) at $\delta=0.07$.

\begin{figure}[b]
	\centering
	\includegraphics[width=1\linewidth]{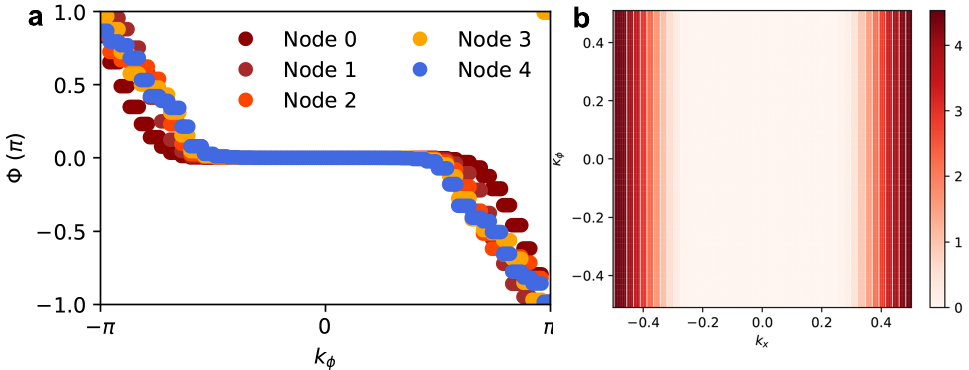}
	\textcolor{black}{\caption{(a) Calculated flat bands Zak phase at different $k_\phi$ in Fig. \ref{fig:largen}, only when $k_\phi=0,\pi$ corresponding a topological $\mathbb{Z}_2$ index $\phi=0,\pi$ due to the inversion symmetry. (Note that $\pm\pi$ are consistent for they represent Wannier center at left/right boundary of unit cell) (b) Berry curvature for Nodes 0 flat band, the Chern number is calculated to be 1.0249799870137477 at $400\times50$ mesh in ($k_x,k_\phi$) space}}5
	\label{fig:topophasen40}
\end{figure}	
\textcolor{black}{
\section{$\Theta$ real-space finite discretization}
We consider constructing a one-dimensional periodic operator \( \Theta \) that includes a spatially varying dielectric constant \( \varepsilon(x) \) and a non-uniform grid distribution. The lattice structure is parameterized by the number grid points \( N_x \) and the displacement parameter \( \delta \). The lattice positions are given by a list of lattice points \( \{x_i\} \) and a spacing vector \( \{s_i\} \) between adjacent points, with boundary conditions ensuring periodicity. The unit (A segment with length $a_{1,2}$) is further divided into two subregions with dielectric constants \( \varepsilon \) and \( \varepsilon_0 \) assigned respectively, and each subregions are discrete with $u$ grid points. For each subregion, the position of the grid points is determined by the parameter \( \kappa \), which controls the distribution within each segment as follows: when \( s < u \), the grid point position is given by
\begin{gather}
	x_{n,s} = x_n + \kappa s_i \frac{s}{u},
\end{gather}
and when \( s \geq u \), the grid point position is
\begin{gather}
	x_{n,s} = x_n + \kappa s_i + (1 - \kappa) s_i \frac{s - u}{u}.
\end{gather}
The operator we construct describes a one-dimensional second-order differential operator with a spatially varying coefficient
\begin{gather}
	\Theta = -\frac{d}{dx} \left( \frac{1}{\varepsilon(x)} \frac{d}{dx} \right),
\end{gather}
which is discretized over a non-uniform grid using finite difference approximations. The matrix elements are computed as follows: the diagonal element is $\Theta_{n,n} = B_{2n} + A_n$, the upper off-diagonal element is $\Theta_{n,n+1} = -(A_n + B_{1n}),$ and the lower off-diagonal element is $\Theta_{n,n-1} = -B_{3n},$ where \( A_n \) and \( B_{mn} \) depend on the dielectric constants and the distances between adjacent grid points. Specifically, \( A_n \) is given by
\begin{gather}
	A_n = \frac{\varepsilon_{n+1}^{-1} - \varepsilon_n^{-1}}{(x_{n+1} - x_n)^2},
\end{gather}
and the terms \( B_{1n} \), \( B_{2n} \), and \( B_{3n} \) are given by
\begin{gather}
	B_{1n} = \frac{\varepsilon_n^{-1}}{(x_n - x_{n-1})(x_{n+1} - x_n)}, \\
	B_{2n} = \frac{\varepsilon_n^{-1}}{x_n - x_{n-1}} \left( \frac{1}{x_n - x_{n-1}} + \frac{1}{x_{n+1} - x_n} \right), \\
	B_{3n} = \frac{\varepsilon_n^{-1}}{(x_n - x_{n-1})^2}.
\end{gather}
Periodic boundary conditions are enforced by phase factors \( \exp(\pm i k_x ) \) to connect the first and last grid points, such that
\begin{gather}
	\Theta_{0,N_x-1} = -B_{3,n=0} e^{-i k_x}, \\
	\Theta_{N_x-1,0} = - (A_{N_x-1} + B_{1,N_x-1}) e^{i k_x }.
\end{gather}
This construction results in a sparse  matrix that captures the spatial inhomogeneity and periodicity of the system, making it suitable for analyzing the electronic properties of one-dimensional dielectric-modulated systems, i.e. breathing lattice. The parameters $\varepsilon$, $\varepsilon_0$, and $\kappa$ provide detailed control over the local dielectric environment and grid density, which is crucial for accurately modeling the microstructural variations of the system and more suitable for topological calculations. As illustrate in Fig. \ref{fig:topophasen40}, by using the eigenvector of real-space discreted $\Theta$ operator, we compute the Zak phase and Berry curvature at $(k,k_\phi)$ space, and $k_\phi\in(0,2\pi]$ is defined as the shift of the grid points \( \{x_i\} \rightarrow \{x_i+k_\phi N a_0/2\pi \}\). 
}
\section{Additional information on Q-factors}
\begin{figure*}
	\centering
	\includegraphics[width=0.8\linewidth]{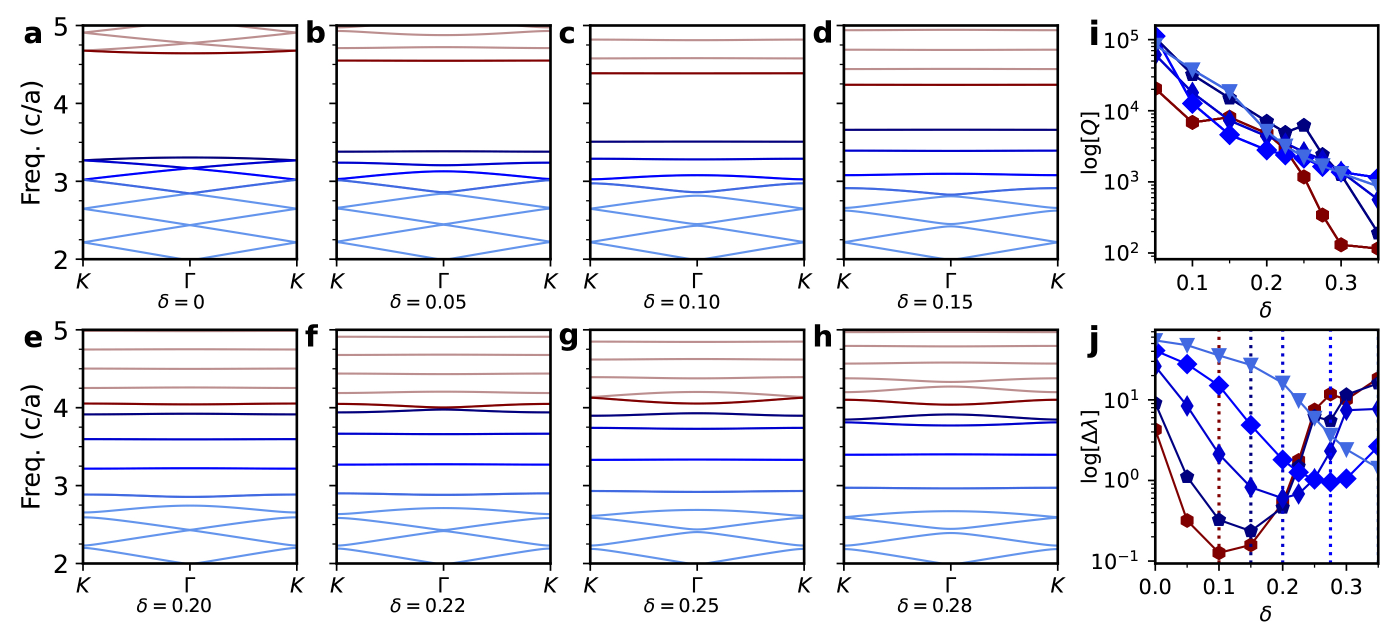}
	\caption{(a-h) present the evolution of flat bands for the TE-polarization, which exhibits similar qualitative behavior as the TM polarization shown in Fig. \ref{main-fig:transision}. The dependence of Q-factors of flat bands on the breathing strength $\delta$  is shown in (i).}
	\label{fig:s1transisionte}
\end{figure*}
\begin{figure}[b]
	\centering
	\includegraphics[width=\linewidth]{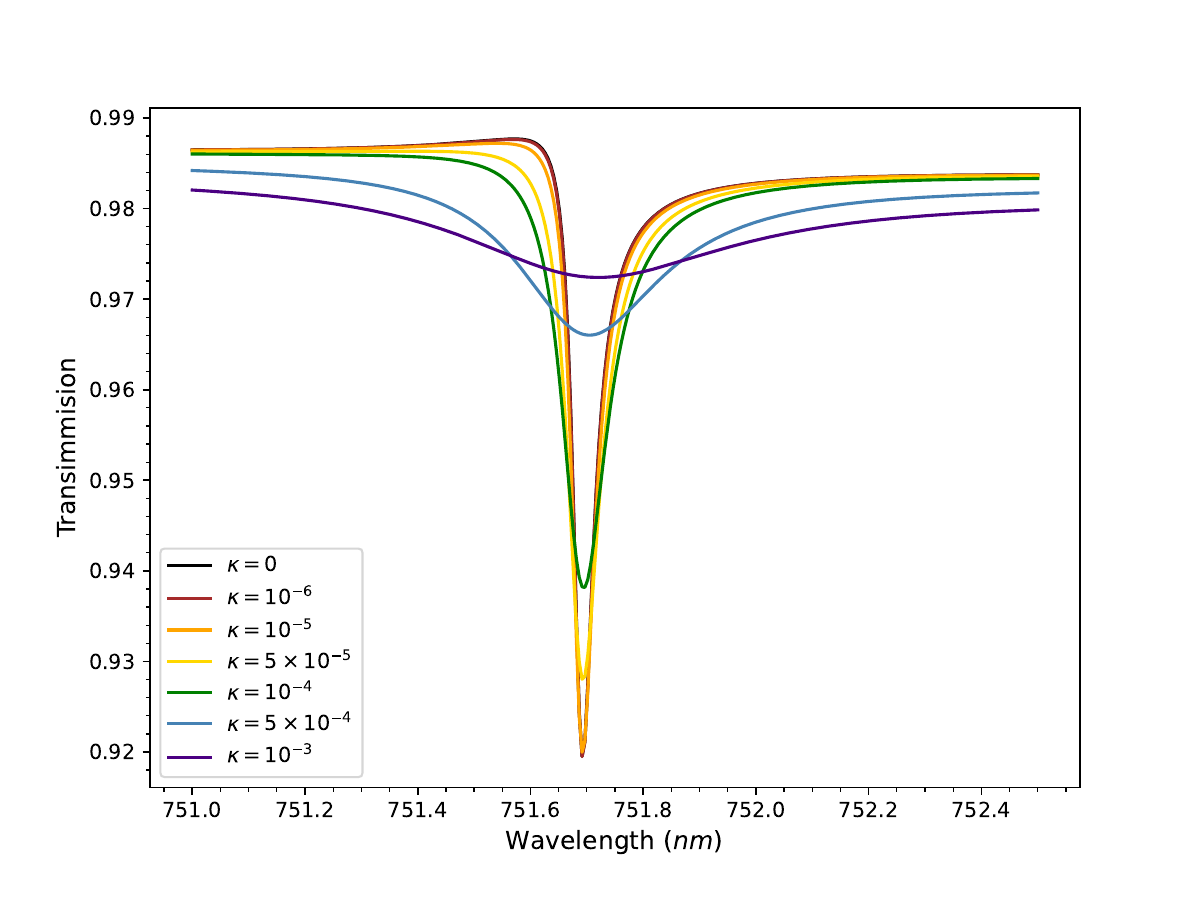}
	\caption{\textcolor{black}{Transmission spectra of photonic crystal slabs of superlattice with different loss $\kappa$ for Fig. \ref{main-fig:bandstructure} (f) .}}
	\label{fig:materialloss}
\end{figure}
\begin{figure}
	\centering
	\includegraphics[width=\linewidth]{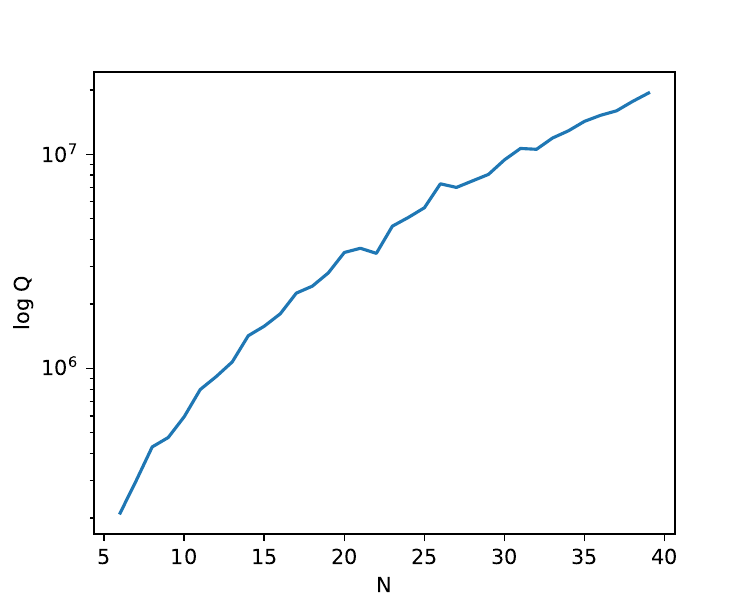}
	\caption{\textcolor{black}{Theoretical Q-factor without substrate with $N$ changing.} }
	\label{fig:qvaringwithn}
\end{figure}

Fig. \ref{fig:s1transisionte}(i) illustrates the dependence of Q-factors on the structural parameter $\delta$ for flat bands. Roughly speaking, the Q-factors of all bands decrease monotonically as $\delta$ increases. \textcolor{black}{As explained in the main text, for the case of $\delta=0$, the dispersive guided modes below the light cone confine light strongly with extremely large Q-factors. When the translational symmetry is broken by slightly perturbing the original structure to a super-lattice, the bands are folded above the light cone, and the Q-factors decrease due to the radiation loss. Moreover, in contrast to the newly formed flat bands from valence bands, where the high Q-factor is found for larger $\delta$, this behavior can be elucidated by considering the competition between the leakage and localization of the photonic modes. For small $\delta$, the localization effect dominates, resulting in stronger localization for earlier generated flat bands since they have lower group velocity $dE/dk$. In contrast, the leakage becomes dominant for larger $\delta$, and the earlier generated flat bands tend to be radiated more rapidly, leading to smaller Q-factors. Additionally, as depicted in Fig. \ref{fig:qvaringwithn}, Q-factors can also enhanced by increasing $N$. Material loss is also considered by introduce the imagery part of refractive index $\kappa$, as depicted in Fig. \ref{fig:materialloss}. The Q-factor will not be affected for the typically $\kappa \le 10^{-5}$ in $\mathrm{TiO}_2$ \cite{refractiveindex_TiO2_Siefke}.}

\begin{figure}
	\centering
	\includegraphics[width=\linewidth]{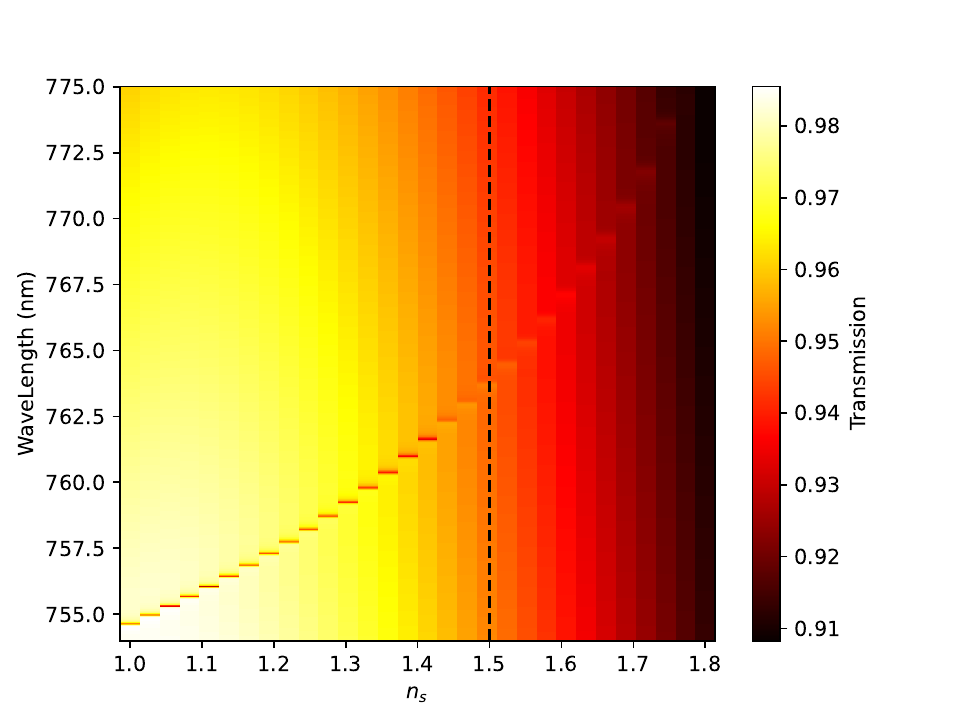}
	\caption{\textcolor{black}{Transmission of  high-Q flat band modes. The dashed black line denotes the refractive index of glass substrate.}}
	\label{fig:substratefpeffect}
\end{figure}

\textcolor{black}{Additionally, in the main text, the theoretical Q value of $1.5\times10^5$ corresponds to the free-standing structure without substrate. However, it is very difficult to fabricate such free-standing membrane, especially considering that our system is a 1D grating structure. In practice, the free-standing structure usually use 2D periodic lattices milled with air holes in the membrane to maintain the samples unbroken. Since the main aim of the experiment in this work is to experimentally confirm the theoretical predictions of flat bands, we only fabricated samples on the substrate. However, our numerical calculations show that the substrate affects the achieved Q-factor strongly. In the Fig. \ref{fig:substratefpeffect}, we present the transmission spectrum for the breathing superlattice by varying the refractive index of the substrate from 1 to 1.8. As shown in the Fig. \ref{fig:substratefpeffect} , the high-Q flat band resonance shifts towards long wavelengths with increasing the substrate refractive index. Meanwhile, due to interface scattering of substrate layer, radiation losses are significantly enhanced, causing the incident energy to scatter into multiple directions. This scattering not only reduces the overall transmission but also leads to a decrease in the Q-factor. As a result, achieving and observing high-Q flat bands in experimental settings becomes challenging.}

\section{eigen-field distributions at different k points}
\begin{figure}[h]
	\centering
	\includegraphics[width=1\linewidth]{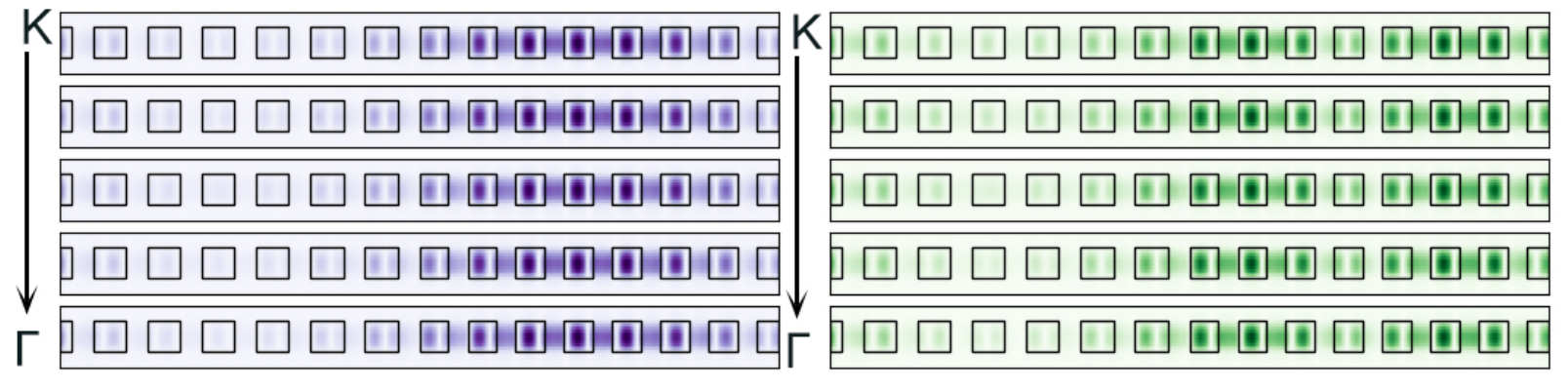}
	\caption{Field localization for two lowest flat bands in Fig. \ref{main-fig:bandstructure}(a,b) along $K$ to $\Gamma$ points with $5$ simples(from up to down), the field is localized everywhere in Brillouin zone. }
	\label{fig:supfield}
\end{figure}

To confirm the localization of light, the eigenfields of the lower two flat bands at five $k$-points from $K$ to $\Gamma$ are plotted in Fig .\ref{fig:supfield}. And it is clearly that the field distribution is highly localized in almost same manner across the Brillouin zone. 
\section{Derivation of the gap opening rate}

The gap opening rate directly reflects the flat band transition at small $\delta$, as illustrated in Figs. \ref{main-fig:transision} and \ref{fig:s1transisionte}. From TB, the structure reduces to the SSH model \cite{PhysRevLett.42.1698} when $N=2$. The energy bands can be expressed as:
$
E_\pm(k)=\pm\sqrt{t_1^2+t_2^2+t_1t_2\cos(q\cdot a)},
$
where $t_1,t_2$ are the hopping energies between the expanded and shrinking regions. At the K point, the energy becomes $E_\pm=\pm|t_1-t_2|$ and the band gap is $2|t_1-t_2|$. We can easily see that the gap closes and forms a Dirac point at $t_1=t_2=t_0$. If $t_{1,2}=t_0(1\pm\delta_{T})$, then we obtain the band gap $\delta E=2 t_0\delta_{T}$.

When we increase $N$, there are $N/2$ Dirac points at the $K$ point due to band folding. Moreover, there are also $N/2-1$ Dirac points at the $\Gamma$ point when $N\geq 2$. As we gradually increase the perturbation parameter $\delta_{T}$, these Dirac points open up cascadingly. TB containing only nearest-neighbor hopping reads:
\begin{gather}
	H(k)=\begin{bmatrix}
		\mu  &t_1& 0 &...&0& t_2 e^{ik}\\
		t_1&\mu  &t_1 &...&0& 0 \\
		0  &t_1 & \mu &...&0& 0 \\
		...&...&...&...&...&...\\
		0  & 0  & 0&...&\mu& t_2\\
		t_2 e^{-ik}&0&0&...&t_2&\mu
	\end{bmatrix}
	_{N\times N},
\end{gather}

\begin{figure}[h]
	\centering
	\includegraphics[width=0.65\linewidth]{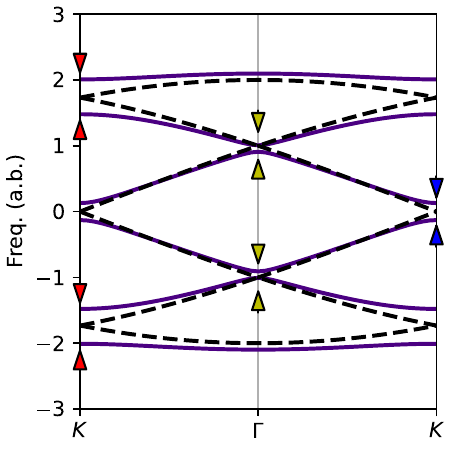}
	\caption{Purple line: $N=6$ band structure with $\delta_T=0.1$. The colored arrows label the gaps. Dashed line: $\delta_T=0$ band without gap opening.}
	\label{fig:s1tbband}
\end{figure}

This model can be analytically solved at higher symmetry points for small $N$. Here we solve it under $N=6$, as depicted in Fig. \ref{fig:s1tbband}. It has three Dirac points at $K$ and two Dirac points at $\Gamma$.  Band Gaps open when  $\delta_T$ is introduced. After Taylor expansion in terms of $\delta_T$, we can obtain the rate of gap opening at $K$ and $\Gamma$, respectively:
\begin{gather}
	\begin{cases}
		\delta E_K^{\mathrm{red}}(\delta_{T})= \frac{8 t_0}{3}\delta_{T}-\frac{40  t_0}{81}\delta_{T}^3+O\left(\delta_{T}^5\right),\\
		\delta E_K^{\mathrm{blue}}(\delta_{T})= \frac{4 t_0}{3}\delta_{T}-\frac{80 t_0}{81}\delta_{T}^3+O\left(\delta_{T}^5\right),\\
		\delta E_\Gamma^{\mathrm{yellow}}(\delta_{T})=\frac{8  t_0}{3}\delta_{T}^2+O\left(\delta_{T}^4\right),
	\end{cases}
\end{gather}
where red, blue and yellow arrows label the gaps in Fig. \ref{fig:s1tbband}. Therefore, as we gradually increase $\delta_T$, gaps at $K$ would open faster than at $\Gamma$, and the outer bands (closer to the gap) open faster, which agrees with the evolution process shown in the main text.
\begin{figure*}
	\centering
	\includegraphics[width=\linewidth]{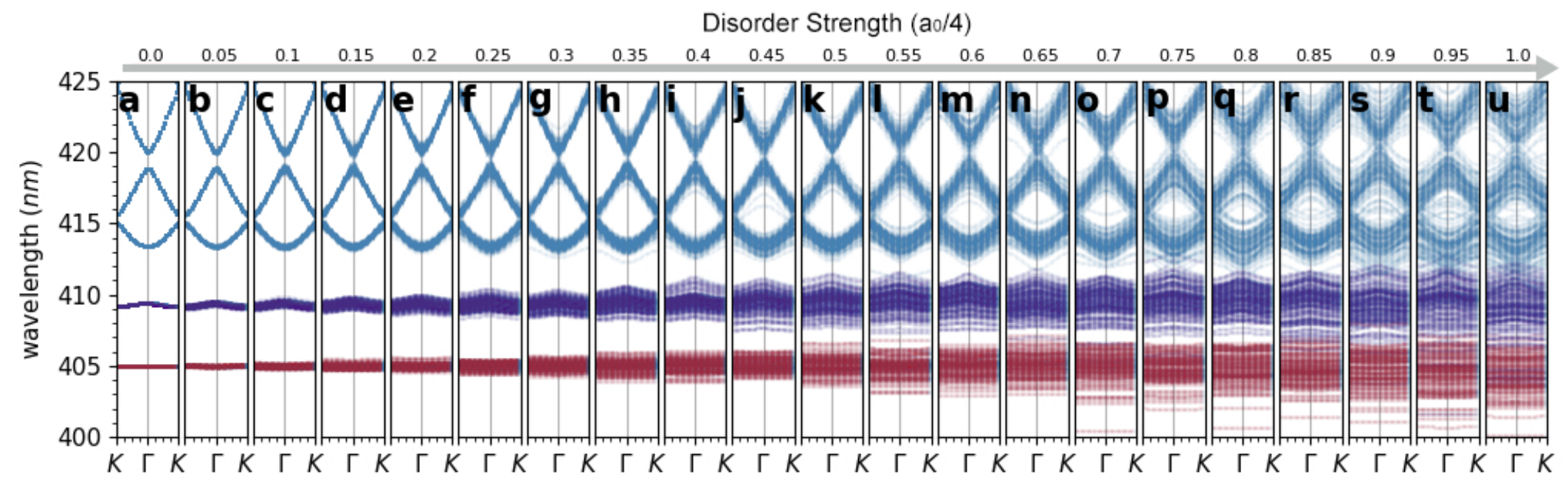}
	\caption{(a-t): Evolution of the band structure for photonic superlattices in experimental parameters when by increasing the intra-cell disorder of the lattice with $21$ disorder’s strength from $0$ to $1$ in the unit of $a_0/4$. The corresponding DOS plot is shown in Fig.  \ref{main-fig:expset}(f). }
	\label{fig:s4intra-cell disorder}
\end{figure*}
\begin{figure}
	\centering
	\includegraphics[width=0.7\linewidth]{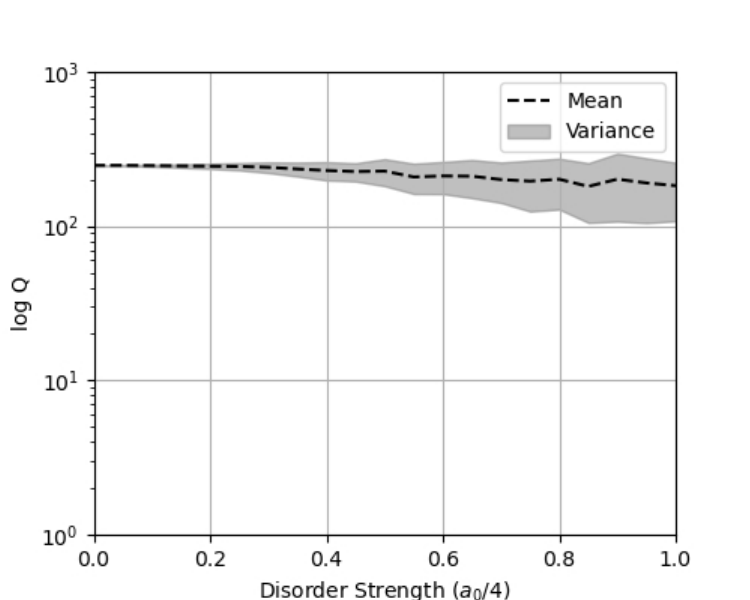}
	\caption{Dependence of the Q factor on the disorder strength. The shading region represents the variance of Q for 100 different samples at a fixed disorder. The dashed curve represents the mean value of Q for 100 samples}
	\label{fig:tedisorderqdos}
\end{figure}
\section{Robustness of flat bands against intra-cell disorder}
	The robustness of flat bands could be understood intuitively from the interference effect of light. As discussed in the last paragraph of Section “\textit{Physical Mechanism}”, the formation of flat bands comes from the construction of new scattering channels between the $i$th and $(i\pm N \pm j)$th sub-bands with $1\le j\ll N$.   Note that the scatterings between the $i$th and $i±N$th subbands give rise to the first band gap of the original simple lattice. These scattering channels represents momentum transfer which corresponds long-range physics at the band edge. Adding weak disorder at short-distance ($\sim a_0$) will not not affect the physics at long distances. Alternatively, above analysis can be intuitively understood from the localization of the eigen-fields in the superlattice. As shown in Figs. \ref{fig:supfield}, the distributions of electric and magnetic fields of flat bands are strongly localized in the expanded and shrunken regions, and the localization of the eigen-fields is not sensitive to local details of each site in the superlattice, since the formation of flat bands require large $N$. Therefore, introducing weak intra-cell disorder can only change the interference condition slightly, thereby do not affect flat bands significantly. 
	The above analysis  has been numerically verified. For the weak disorder that doesn’t close the band gap, the flat bands remain featuring very narrow bandwidths as confirmed Fig. by \ref{fig:s4intra-cell disorder}, where the lattice site positions are randomly displaced from regular lattice site. Increasing intra-cell disorder implies a wider range of possible displacements from the original positions of the sites. The maximum range is  $a_0/4$ (Over $a_0/4$ may lead to the overlap of strips). As shown in Fig. \ref{fig:s4intra-cell disorder}, most flat bands survive even in the strong disorder, and all the flat bands survive in the weak disorder,  demonstrating the robustness under certain intra-cell disorder. In Figs. \ref{main-fig:expset} (f) and \ref{fig:s4intra-cell disorder}, we compute 100 time for each fixing strength of disorder, stronger disorders inevitably leading larger uncertainty of position of flat bands. 

	We also numerically calculated the dependence of the Q factor on the intra-cell disorder for photonic crystal slab with experimental parameters and the disorder strength varying from $0$ to $a_0/4$. As demonstrated in the  Fig. \ref{fig:tedisorderqdos}, the averaged Q factors of flat bands are robust for weak disorder when the disorder strength is smaller than $a_0/8$. As the disorder increases further, the average Q factor decreases while its variance increases, indicating a stronger dependence on the disorder strength.  However, the achieved Q values remain in the order of $100$.
\begin{table*}[htbp]
	\centering
	\begin{tabular}{cccccccccccc}
		\hline
		\hline
		Range of $\gamma$ & $c_0^{(0)}$ & $c_1^{(0)}$ & $c_2^{(0)}$ & $c_0^{(2)}$ & $c_1^{(2)}$ & $c_2^{(2)}$ & $c_3^{(2)}$ & $c_0^{(4)}$ & $c_1^{(4)}$ & $c_2^{(4)}$ & $c_3^{(4)}$ \\
		\hline
		$0.25 < \gamma < 0.5$ & $-0.06$ & $0.51$ & $0.76$ & $-34.6$ & $116.8$ & $889.4$ & $-8591.0$ & $-106187.1$ & $426297.2$ & $-459244.0$ & $-24651.9$ \\
		$0.5 < \gamma < 1.0$ & $-0.20$ & $1.03$ & $0.24$ & $-25.9$ & $96.0$ & $-58.3$ & $-309.8$ & $-117.5$ & $597.6$ & $-443.9$ & $521.6$ \\
		$1.0 < \gamma < 1.5$ & $-0.33$ & $1.31$ & $0.09$ & $1.8$ & $-3.6$ & $1.5$ & $6.5$ & $23.0$ & $-43.0$ & $18.2$ & $-19.6$ \\
		$1.5 < \gamma < 2.0$ & $-0.42$ & $1.43$ & $0.05$ & $1.8$ & $-2.3$ & $0.7$ & $4.5$ & $636.5$ & $-801.6$ & $239.2$ & $-408.2$ \\
		$2.0 < \gamma < 2.5$ & $-0.49$ & $1.50$ & $0.04$ & $-106.6$ & $102.8$ & $-23.8$ & $-189.4$ & $-289.1$ & $271.3$ & $-62.1$ & $134.1$ \\
		$2.5 < \gamma < 3.0$ & $-0.55$ & $1.55$ & $0.03$ & $11.0$ & $-8.3$ & $1.6$ & $13.8$ & $-148.1$ & $111.0$ & $-20.5$ & $50.0$ \\
		$3.0 < \gamma < 3.5$ & $-0.61$ & $1.58$ & $0.02$ & $76.5$ & $-48.1$ & $7.5$ & $71.8$ & $-197.9$ & $123.8$ & $-19.3$ & $51.0$ \\
		$3.5 < \gamma < 4.0$ & $-0.66$ & $1.61$ & $0.02$ & $36.8$ & $-19.9$ & $2.7$ & $27.3$ & $87.4$ & $-47.1$ & $6.3$ & $-17.8$ \\
		$4.0 < \gamma < 4.5$ & $-0.71$ & $1.64$ & $0.01$ & $16.2$ & $-7.6$ & $0.9$ & $9.8$ & $-90.1$ & $42.7$ & $-5.1$ & $14.8$ \\
		\hline
		\hline
	\end{tabular}
	\caption{Fitted Coefficients $c_n^{(p)}$s for different ranges of $\gamma$ with respective range of $\gamma$ in Eq. (\ref{eq:ana}). When $\gamma<0.25$ one can only use the second order perturbation theory Eq. (\ref{eq:pertubation theroy})  }
	\label{tab:data}
\end{table*}

\section{Analytic dispersion relation from the continuum model}
Here we show the analytic result in the main text with the CM model Eq. (\ref{main-eq:CM_mat}). This provides an approximate dispersion relation by considering only the nearest band coupling.
To obtain the band dispersion, we first consider $\gamma\leq\left(\frac{1}{2}\right)^2$.We express the Hamiltonian Eq. (\ref{main-eq:CM_mat}) as $H = H^{(0)} + V$, where $\langle m | H^{(0)} | n \rangle = (q + n)^2 \delta_{m, n} - q^2 \delta_{0, 0}$, and $\langle m | V | n \rangle = q^2 \delta_{0, 0} + \gamma (\delta_{m, n + 1} + \delta_{m, n - 1})$. By using second-order perturbation theory to the uppermost band ($n = 0$), we calculate the energy correction:
\begin{gather}
	\Delta E_0^{(2)}(q)=\gamma^k\sum_{Q\neq 0}\frac{(\delta_{0,Q+1}+\delta_{0,Q-1})(\delta_{Q,1}+\delta_{Q,-1})}{(E_0-E_Q)},
	\label{eq:pertubation theroy}
\end{gather}
where $E_0=0$, $E_Q=(q+Q)^2$ if that $|E_0-E_Q|> \gamma$ is satisfied. Then, we have
$
	\epsilon_0(q)\approx q^2-\gamma^2\left(\frac{1}{(q+1)^2}+\frac{1}{(q-1)^2}\right)
	\approx -2 \gamma ^2+\left(1-6 \gamma ^2\right) q^2+O\left(q^4\right)=F_0+F_2q^2+O\left(q^4\right).
$
This formula holds well until $\gamma=\left(\frac{1}{2}\right)^2$, with the coefficient $F_2$ continuously decreasing as $\gamma$ increases. Beyond this point, band gaps will emerge at the $K$ point. Furthermore, near $\gamma=1$, where $H^{(0)}_{1,1}=(q+1)^2$ and $H^{(0)}_{-1,-1}=(q-1)^2$ equal $\gamma$ at $q=0$, a new gap opens at the $\Gamma$ point. We can find that the gap at $K$ always opens near $\gamma=(\frac {2n+1} 2)^2$ and the gap at $\Gamma$ opens near $\gamma=n^2$. 
When $\gamma>\left(\frac{1}{2}\right)^2$, we can diagonalize $H_1$ and 
$$
\tilde H_2=
\begin{bmatrix}
	&(q-2)^2&\gamma &&\\
	&\gamma &(q-1)^2&\gamma\\
	&&\gamma &q^2&\gamma\\
	&&&\gamma &(q+ 1)^2
\end{bmatrix},
$$ which are analytically obtained in \texttt{mathmatica} ($\tilde H_2$ only used for $q>1/2$. And, $q<1/2$ can be obtained by using the inversion symmetry). We then expand the energy band terms in powers of $q$, approximating it as follows:
\begin{gather}
	\epsilon_0(q)\approx\sum_{n=0}^2 c^{(0)}_n \gamma^n +\frac{\sum_{n=0}^2 c^{(2)}_n \gamma^n}{\sum_{n=0}^3 c^{(2)}_n \gamma^n} q^2 +\frac{\sum_{n=0}^2 c^{(4)}_n \gamma^n}{\sum_{n=0}^3 c^{(4)}_n \gamma^n} q^4
	\label{eq:analytica expression}
\end{gather} 
This approximation is valid for $\left(\frac{1}{2}\right)^2 < \gamma$. The parameters $c_n^{(p)}$ are obtained through numerical fitting. Once $c_n^{(p)}$ are determined, the explicit expressions for these parameters can be obtained across a broad range of $\gamma$. We auto-fit these parameters using the \texttt{curvefit} function of the \texttt{scipy} package, employing a step size of $0.5$. In particular, the coefficient of $q^2$ can be re-expressed as:
\begin{gather}
	F_m(\gamma)=
	\begin{cases}
		\left[1+\frac{c_3^{(m)}\gamma^3}{c_0^{(m)}+c_1^{(m)}\gamma+c_2^{(m)}\gamma^2}\right]^{-1},\quad \gamma>\left(\frac 12\right)^2,\\
		1-6\gamma^2, ~~\quad\quad\quad\quad\quad\quad\quad\quad~~ \gamma\le \left(\frac 12\right)^2,
	\end{cases}		
	\label{eq:ana}
\end{gather} 
which is given in main text. And, it is obviously that $\lim_{\gamma\rightarrow \infty} F_m=0$. The fitted coefficients are listed in Tab. \ref{tab:data}.

\section{Renormalization of $\gamma$ by scaling N }

The dependence of flat bands on $N$ can be understood as $N$ scaling the system, which effectively renormalizes the scattering strength $\gamma$. (Note that the dependence of the number of flat bands on $\gamma$ has been investigated in Eq. (\ref{main-eq:CM_mat}) in the manuscript.) Based on the analysis in the section "\textit{Microscopic mechanism}" of the main text, the number of flat bands can be understood by comparing the renormalized scattering strength $\gamma'(N,\gamma)$ (which corresponds to the off-diagonal elements) and the diagonal elements of the CM Hamiltonian. For example, if $\gamma'(N,\gamma)$ exceeds $M$ diagonal elements (where $M$ is an integer), the structural breathing will cause $M$ sub-bands to be flattened, i.e., generate $M$ flat bands.

The detailed  process  of how $N$ renormalize $\gamma$ to $\gamma'(N,\gamma)$ is represented as follows. From Eq. (\ref{main-eq:CM_mat}) in the main text:
\begin{gather}
	H_m(q)=\begin{bmatrix}
		(q-m q_0)^2&\gamma I_{m-1}&0\\
		\gamma I_{m-1}^T&H_{m-1}(q)&\gamma I_{m-1}^T\\
		0&\gamma I_{m-1}&(q+m q_0)^2
	\end{bmatrix},
	\label{eq:CM_mat}
	\nonumber
\end{gather}
where $q = q_0q' \in q_0(-1/2,1/2]$ and $\gamma$ is the inter-band scattering strength. Naturally, $N$ can be represented by $q_0^{-1}$ , which is now a scaling parameter. By taking the term of $q_0$ out from the matrix, $H_m$ can be written as the following form:

\begin{gather}
	H_m(q', q_0, \gamma) = q_0^2 \left[
	\begin{matrix}
		(q'-m)^2 & \frac{\gamma}{q_0^2} I_{m-1} & 0 \\
		\frac{\gamma}{q_0^2} I_{m-1}^T & H_{m-1}(q') & \frac{\gamma}{q_0^2} I_{m-1}^T \\
		0 & \frac{\gamma}{q_0^2} I_{m-1} & (q'+m)^2
	\end{matrix}
	\right]\nonumber \\
	= \left(\frac{2\pi}{N a_0}\right)^2 H_m\left(q', 1, \frac{(N a_0)^2 \gamma}{4\pi^2}\right),
\end{gather}

where $N a_0$ is the length of the unit cell, and $I_m = [1, 0, \ldots, 0]$ is an $m$-dimensional vector as defined in the main text. The above equation indicates that the target Hamiltonian, which depends on  $q_0$, can be represented by $H_m(q', 1, \gamma'(N,\gamma))$ and
\begin{gather}
	\gamma'(N,\gamma)= \frac{\gamma }{q_0^2}=\frac{(N a_0)^2 \gamma}{4\pi^2}
\end{gather}
where $\gamma'>\gamma$ for $N$ is large. If we denote the spectrum of $H_m (q',1,\gamma)$ as $\varepsilon_m (q',\gamma)$, then the spectrum of $H_{m,N} (q',\gamma):=H_m (q',2\pi/(Na_0 ),\gamma)$ can be written as

\begin{gather}
E_{m,N}(q', \gamma) = \left(\frac{2\pi}{N a_0}\right)^2 \varepsilon_m\left(q',\gamma' (N,\gamma)\right).
\label{eq:renormalize}
\end{gather}
Therefore, once we obtain the band structure data of a CM model at all possible $\gamma$ numerically, we can scale to find all the information on different $N$ based on help of Eq. (\ref{eq:renormalize}). Consequently, the dependence of the number of flat bands on N can be transformed to the dependence of the number of flat bands on $ \gamma'$, which have been discussed in the main text.

We can also investigate other properties of flat bands through this process. For example, if we consider the empirical dispersion formula of $E_q (0)$ similar to $\epsilon_0(q)$ in Eq. (\ref{eq:analytica expression}), one arrive at
\begin{gather}
E_{N,q'} (0,\gamma)=\left(\frac{2\pi}{Na_0 }\right)^2 (F_0 (\gamma' )+F_2  (\gamma' ) q'^2+F_4 (\gamma') q'^4 )\nonumber\\
=\left(\frac{2\pi}{Na_0 }\right)^2 E_{q'} (0,\gamma')
\end{gather}
where $0$ labels the upper most flat band, and $\gamma'=\gamma/q_0^2=(Na_0 )^2 \gamma/4\pi^2>\gamma$.

\section{ Experimental measurement and data post-processing }

The samples shown in Fig.  \ref{main-fig:exptransision} in the main text are fabricated with parameters tabled in Tab. \ref{tab.parameter}. Due to fabrication inaccuracies, we have two types of $\delta$, neither of which affects the physics of the flat bands. Specifically, the parameters are defined as $a_{1,2}=a_0(1\pm\delta_1)$ and $d_{1,2}=\kappa a_0(1\pm\delta_2)$. In the calculations, we find that the larger of $\delta_{1,2}$ dominates the band structure.
\begin{table}[h]
	\centering
	\begin{ruledtabular}
		\begin{tabular}{ccccc}
			Sample & $\delta_1$ & $\kappa$ & $\delta_2$ & $a_0$ ($nm$) \\
			\colrule
			a & 0.024 & 0.606 & 0.039 & 191.25 \\
			b & 0.08  & 0.662 & 0.07  & 200.25 \\
			c & 0.109 & 0.594 & 0.066 & 201.4  \\                                                                                                                                                                                                                                                                                                                                                                                                                                                                                                                                                                                                                                                                                                                                                                                        
		\end{tabular}
	\end{ruledtabular}
	\caption{Experimental measures parameters of samples in Fig. \ref{main-fig:expset} (h-j).}
	\label{tab.parameter}
\end{table}
\begin{figure}
	\centering
	\includegraphics[width=0.95\linewidth]{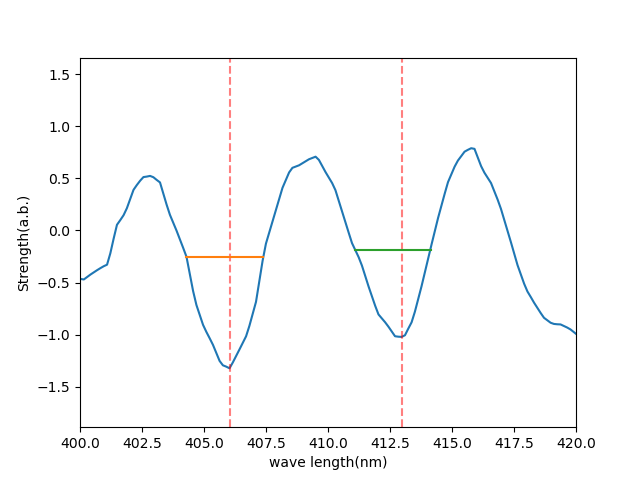}
	\caption{Typical transmission spectrum from experimental at fixed incident angle. Full width at half maximum (FWHM) of peaks are highlighted with colored line. }
	\label{fig:s10spec}
\end{figure}
The measurement results shown in Fig. \ref{main-fig:expset} and \ref{main-fig:exptransision} are post-processed from the original transmission spectrum $T(\lambda,\phi)$ when the phase matching condition:
\begin{gather}
	p=\frac{2\pi}{\lambda}\sin(\phi),
\end{gather}  
where $\phi$ is the incident angle. In measurement, the angle $\phi$ satisfies $-20^\circ<\phi<20^\circ$. The Bloch momentum $q$, which relates with $p$ through band folding, is given by:
\begin{gather}
	p=q + \frac{2\pi m}{a},
\end{gather}  
where $a=Na_0$ is the lattice constant, $m$ is the integer. Through the \texttt{graddata} function in the \texttt{scipy} package, we discretely sample the experimental data $T(\lambda,\phi)$ on a two-dimensional grid, transform the data to $T(\lambda,q)$, and finally obtain the spectrum. 

In the Fig. \ref{main-fig:expset} (g) we measures the FWHM form experimental data and obtain the information corresponding to the imagery eigenvalue, this process can  be clearly found in Fig. \ref{fig:s10spec}, where the colored line illustrate FWHM. This process can be automatic done by using \texttt{find\_peaks} and \texttt{peak\_widths} functions in \texttt{scipy.signal} package. In experimental, we obtain Q factor $\sim100$ based on FWHM extract from transmission spectrum (Fig. \ref{main-fig:expset} (g) and Fig. \ref{fig:s10spec}) by $Q=\lambda_0/\mathrm{FWHM}=\Re{(E)}/\Im{(E)}$, with $\lambda_0$ is the spectral position of the peak.
\begin{figure}
	\centering
	\includegraphics[width=\linewidth]{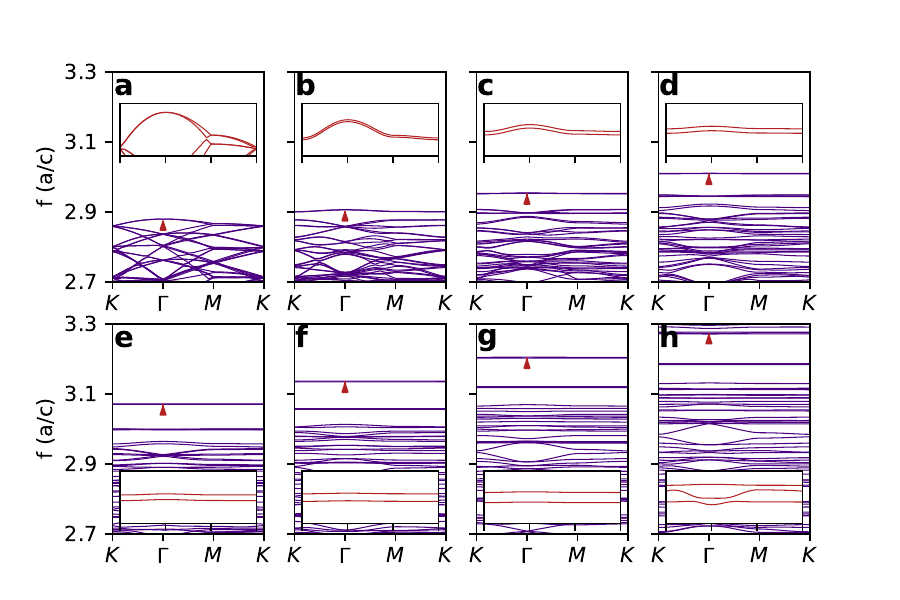}
	\caption{Evolution of flat bands for $\delta= 0, 0.2, · · · , 1.4$ lattice structure is dipicted in main text Fig. 6.Insert: the upper most two flat bands of VBM}
	\label{fig:s5bandtransition}
\end{figure}
\textcolor{black}{\section{2D flat bands}
\subsection{Construction of 2D breathing lattices}
For the partial flat bands lattice, the construction is the same as 1D cases, where the hole radius is set to be $r=0.365(1\mp\delta) a_0$.}

\textcolor{black}{For the 2D triangular superlattice, the lattice vectors are defined as $A_1 = N(2a_2 - a_1)$ and $A_2 = N(a_1 + a_2)$, where $a_1 = a_0[\sqrt{3}, 1]^T / 2$, $a_2 = a_0[0, 1]^T$, and $N = 8$. In Fig. 6(b), $a=|A_i|$ is the lattice constant. The shift parameter $\delta$ divides the unit cell into four regions: $S_{++} : (m < 1/2, n < 1/2)$, $S_{+-} : (m \geq 1/2, n < 1/2)$, $S_{-+} : (m < 1/2, n \geq 1/2)$, and $S_{--} : (m \geq 1/2, n \geq 1/2)$, where $(m, n)$ are the fractional coordinates of the lattice sites with respect to $A_1$ and $A_2$. For the $S_{\pm\pm}$ regions, the lattice sites are generated as $p = i(1 \pm \delta)a_1 + j(1 \pm \delta)a_2$, where $i, j \in \mathbb{Z}$. The radius of the hole is specified as $0.365(1 + \delta)a_0$ in the $S_{++}$ region and $0.365(1 - \delta)a_0$ in other regions. 
\subsection{Variation of 2D flat bands}
In the main text, Fig. 3 displays the band transition in one-dimensional cases, and Fig. 6 presents the omnidirectional flat band achieved using the GME method. Here, we examine the transition of this 2D band structure. Initially, Fig. \ref{fig:s5bandtransition}(a) shows two degenerate bands in the Valence band maximum (VBM), highlighted in red. As $\delta$ increases, these two bands undergo a transition, resembling the fattening observed in the 1D cases. Moreover, the spacing between the two bands widens with increasing  $\delta$.
Subsequently, another group of flat bands, consisting of four bands, exhibits a more intricate evolution compared to the 1D cases. The gap initially opens at the $K$ points, as illustrated in Figs. \ref{fig:s5bandtransition}(a) and (b), while remaining gapless at the $\Gamma$ points. Upon further increasing $\delta$,
the gap also emerges at the $\Gamma$ points, resulting in two groups of bands, as depicted in Figs. \ref{fig:s5bandtransition}(b) and
(c). Following this, the group of flat bands widens, as shown in Fig. \ref{fig:s5bandtransition}(d). With further increasing of $\delta$, as shown in Figs. \ref{fig:s5bandtransition}(e) and (f)
we can find there are $8$ bands under flat bands and separated from others. However, their can be flatten but can not clearly separated from each other as shown in Fig. \ref{fig:s5bandtransition}(g) and
(h). In (h) we can find re-dispersion of two uppermost flat bands from VBM.}
\bibliography{Photonic_flat_band.bib}